\documentclass[11pt,a4paper]{article}
\pdfoutput=1
\usepackage{jheppub}
\usepackage[utf8]{inputenc}
\usepackage{amsmath}
\usepackage{xcolor}
\usepackage{graphicx}
\usepackage{array}
\usepackage{dsfont}
\usepackage{subfigure}

\title{
Dispersion relations in non-relativistic two-dimensional materials from quasinormal modes in Ho\v{r}ava Gravity
}

\author[]{Markus Garbiso, Matthias Kaminski}
\affiliation[]{Department of Physics and Astronomy, University of Alabama,\\ Tuscaloosa, AL 35487, USA}
\emailAdd{magarbiso@crimson.ua.edu}
\emailAdd{mski@ua.edu}

\date{\today}
\begin{document}

\abstract{
We compute dispersion relations of non-hydrodynamic and hydrodynamic modes in a non-relativistic strongly coupled two-dimensional quantum field theory. 
This is achieved by numerically computing quasinormal modes (QNMs) of a particular analytically known black brane solution to 3+1-dimensional Ho\v{r}ava Gravity.  
Ho\v{r}ava Gravity is distinguished from Einstein Gravity by the presence of a scalar field, termed the khronon, defining a preferred time-foliation. 
Surprisingly, for this black brane solution, the khronon fluctuation numerically decouples from all others, having its own set of purely imaginary eigenfrequencies, for which we provide an analytic expression. 
All other Ho\v{r}ava Gravity QNMs 
are expressed analytically in terms of QNMs of Einstein Gravity, in units involving the khronon coupling constants and various horizons. 
Our numerical computation reproduces the analytically known momentum diffusion mode, and extends the analytic expression for the sound modes to a wide range of khronon coupling values. 
In the eikonal limit (large momentum limit), the analytically known dispersion of QNM frequencies with the momentum is reproduced by our numerics. 
We provide a parametrization of all QNM frequencies to fourth order in the momentum. 
We demonstrate perturbative stability in a wide range of coupling constants and momenta. 
}

\maketitle

\section{Introduction}
Right after the discovery of the gauge/gravity correspondence~\cite{Maldacena:1997re}, it was applied in parallel with hydrodynamics to describe strongly coupled fluids~\cite{Policastro2002,
Policastro:2001yc,Bhattacharyya:2008jc}, which are relativistic. 
Most experiments, however, are conducted in non-relativistic materials, some of which may be described best as a non-relativistic strongly coupled fluid.  

A holographic correspondence has been conjectured between Ho\v{r}ava Gravity~\cite{Horava:2009uw,Horava:2009vy,Anderson:2011bj,horava2009} and non-relativistic quantum field theories~\cite{Janiszewski:2012nf,Janiszewski2013}. 
%
An analytical Ho\v{r}ava black brane solution was found~\cite{Janiszewski2015}, and its fluctuations were studied analytically in the hydrodynamic limit~\cite{Davison2016}. A hydrodynamic momentum diffusion mode was found via a field redefinition in what we will refer to as {\it axial sector} of the theory. That field redefinition maps the axial sector of Ho\v{r}ava Gravity to the corresponding sector of Einstein Gravity. However, in the other sector, the {\it polar sector}, this map could only be performed in the special case of setting one Ho\v{r}ava coupling constant to zero, $\lambda=0$. Two hydrodynamic sound modes were found as expected. But this failure of the map at $\lambda\neq 0$ motivates a closer study of the polar sector. 
%

For our purposes, Ho\v{r}ava Gravity is Einstein Gravity with the addition of a scalar field, the {\it khronon}. This scalar field defines a preferred time-foliation and thus breaks Lorentz invariance. 
%
Generally, non-relativistic theories exhibit {\it Lifshitz scaling}, defined as different scaling in time, $t$, and spatial, $x^I$ coordinates, i.e. $t \to \kappa^z t , \, x^I \to \kappa x^I $, with constant $\kappa$ and dynamical exponent $z$. The analytic Ho\v{r}ava black brane solution is peculiar in that it has $z=1$. While this solution still breaks Lorentz boost invariance, and the theory is non-relativistic, see discussion in Sec.~\ref{horSec}, its Lifshitz scaling coefficient $z=1$ is that of a relativistic theory. 
Furthermore, in the case of field theories with Lifshitz scaling $z$ and dimension $d$ it has been claimed that these contain only overdamped eigenmodes, i.e. quasinormal modes on the imaginary frequency axis, in their spectrum if $d\le z+1$ at vanishing momentum~\cite{Sybesma2015,Gursoy:2016tgf}. The analytic Ho\v{r}ava black hole solution has $z=1$, $d=3$ and should thus not be overdamped, should it follow the behavior described in~\cite{Sybesma2015,Gursoy:2016tgf}. These references~\cite{Sybesma2015,Gursoy:2016tgf} consider a different system though, namely a probe scalar in a background with Lifshitz scaling. Therefore it is not clear if also quasinormal modes in Ho\v{r}ava Gravity should show this behavior. The peculiar value of $z=1$, imposed by the one and only analytic Ho\v{r}ava black brane background available to us from the solutions described in~\cite{Janiszewski:2015ura}\footnote{These black brane solutions are smoothly connected to solutions with all possible two-derivative Ho\v{r}ava couplings non-zero~\cite{Janiszewski:2015ura}.}, motivates us to investigate the quasinormal modes of this solution in order to see if there are overdamped modes and to investigate the dispersion relations of this theory. It is a priori not clear, if they should be relativistic or non-relativistic. 

In Ho\v{r}ava Gravity, fields can travel at distinct speeds that can even be infinitely large~\cite{Janiszewski2015}. These speeds are set by the Ho\v{r}ava coupling constants. The existence of distinct speeds implies that fields experience different horizons, i.e. last points of return, in the presence of a black brane. In~\cite{Davison2016} it was conjectured that the relevant horizon for each set of fields is the sound horizon, which is identified as a regular singular point of the set of equations of motion of those fields. For the axial sector the relevant horizon was shown to be the horizon of the spin-2 graviton. In this work we confirm that in the polar sector the relevant horizon is the spin-0 graviton horizon, which here is identical to the universal horizon $r_h$. The universal horizon determines the temperature in the dual field theory~\cite{Janiszewski2015}.

Hence, the main goal of this work is to calculate all quasinormal modes in the axial and polar sectors for a wide range of Ho\v{r}ava coupling values, mode numbers and momenta. From this, we extract the dispersion relation of each mode.  
The questions mentioned above will be answered in passing. 
%
In the polar sector, see Sec.~\ref{sec:AE}, the fluctuation equations for Ho\v{r}ava fields are very complicated. Hence, we use the equivalence between Ho\v{r}ava Gravity and Einstein-\AE ther Theory when the scalar field, the khronon, is required to be hypersurface-orthogonal~\cite{BHATTACHARYYA2013,Jacobson2000}. In the axial sector we show very good numerical agreement of quasinormal modes from both theories. 
%
Compared to Einstein Gravity, in Ho\v{r}ava Gravity, an additional set of quasinormal modes is to be expected, namely those contributed by the khronon. These {\it khronon modes} turn out to be very special and we discuss them in detail in Sec.~\ref{sec:khrononModes}. 

\begin{figure}[h!]
\centering
    \begin{subfigure}
        \centering
        \includegraphics[height=3.1in]{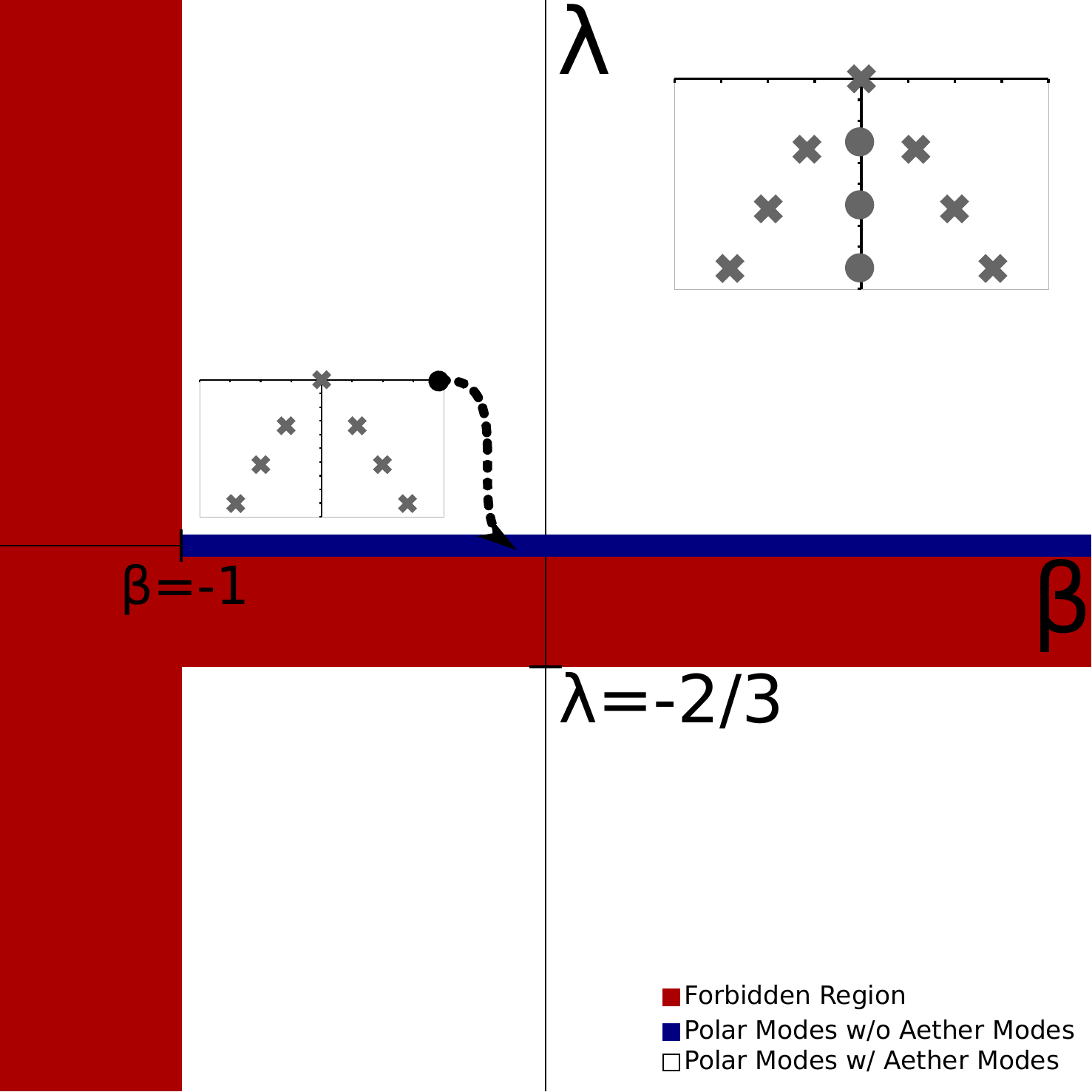}
        \caption{A ($\lambda$, $\beta$)-parameter plot for the {\it polar} sector. The origin of the graph indicates $\lambda = 0$ and $\beta = 0$. The red parameter regions are forbidden~\cite{Griffin2013}. The insets show typical locations of QNMs in the complex frequency plane at vanishing momentum, depending on the values of $\lambda$ and $\beta$. The crosses indicate QNMs caused by metric fluctuations, $h$. Dots indicate {\it khronon modes} caused by fluctuations of the additional ``non-relativistic'' degree of freedom, {i.e.}~the khronon field, $\phi$, in Einstein-\AE ther theory, 
        or equivalently the time-like vector, $u$, in Ho\v{r}ava Gravity. For coupling constant values $\lambda = 0$ and $\beta\ge -1$~(blue line), there are no QNMs associated with the khronon.}
\label{fig:ParameterPolarPlots}
    \end{subfigure}
    ~
    \begin{subfigure}
        \centering
        \includegraphics[height=3.1in]{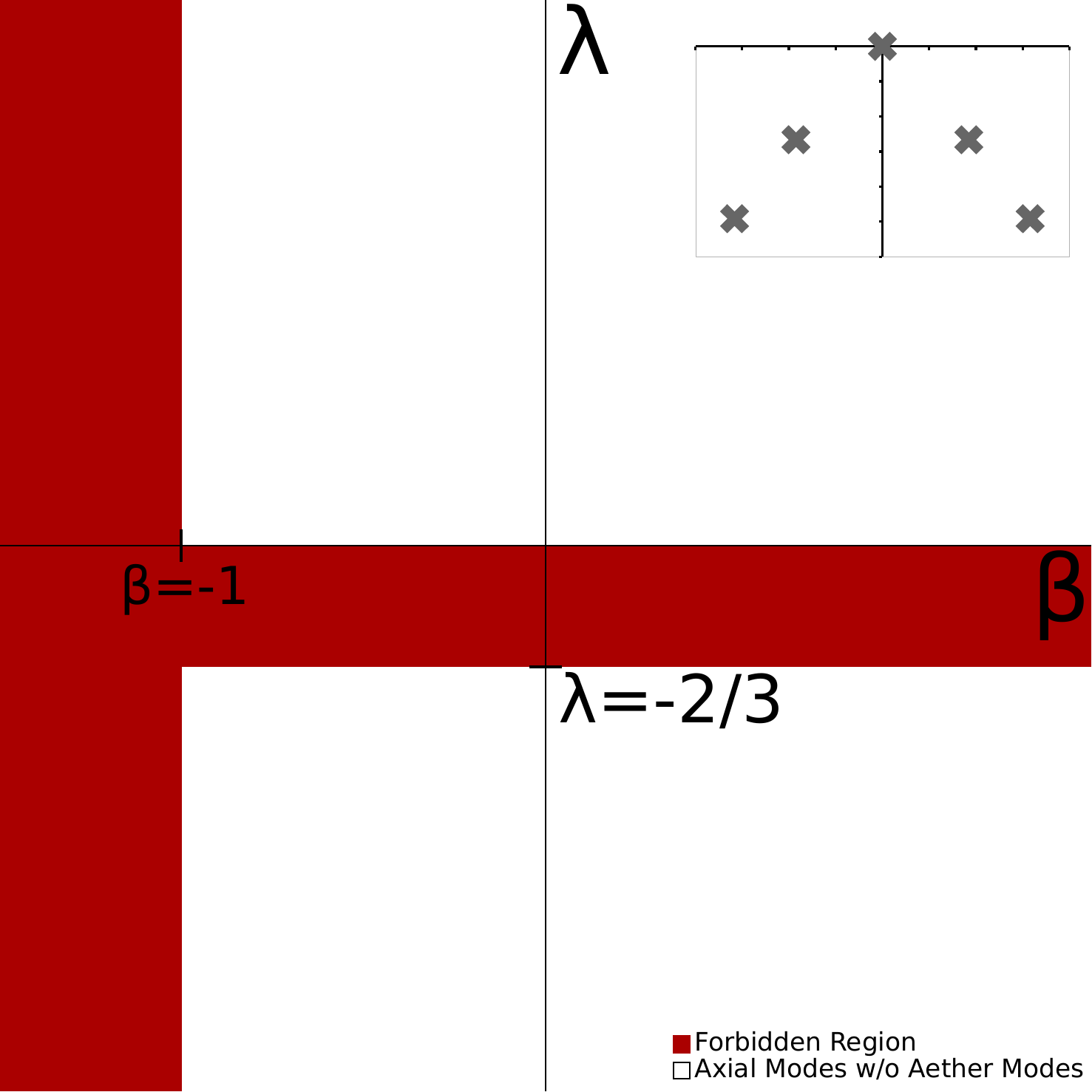}
        \caption{Like Fig.~\ref{fig:ParameterPolarPlots}, but now for the {\it axial} sector. There are no QNMs associated with the khronon~(Einstein-{\AE}ther field), or equivalently with the timelike Ho\v{r}ava Gravity vector field $u$.}
\label{fig:ParameterAxialPlots}
    \end{subfigure}
\end{figure}

We study long-lived modes determining the behavior of the system at late times. Two types of modes are long-lived because their damping is small: hydrodynamic modes with small frequency and momentum, and non-hydrodynamic modes with large momentum. 
Analytic results in the large momentum (eikonal) limit~\cite{Festuccia:2008zx,Morgan:2009vg,Fuini2016}, are a  useful tool to check numerical accuracy and the structure of the quasinormal mode spectrum, see Sec.~\ref{sec:largeMomentumLimit}. 
Our results are discussed in Sec.~\ref{sec:horavaAxialQNMs} and~\ref{secQNMPseudoSpectral}, but visually summarized in Fig.~\ref{fig:ParameterPolarPlots} and Fig.~\ref{fig:ParameterAxialPlots}. The parameter space of Ho\v{r}ava coupling constants $\lambda$ and $\beta$ is shown. The third coupling is set to zero in this work, $\alpha=0$. 
A summary of results and our conclusions are found in Sec.~\ref{sec:conclusions}. Equations of motion, and QNM data is collected in four ancillary files.

The two-dimensional materials we have in mind for application of our results could be thin films (e.g. NbSi~\cite{Hartnoll:2007ih}) or layered structures such as those found in high critical temperature cuprates, see e.g.~\cite{Hartnoll:2007ih,Blauvelt:2017koq} and more generally~\cite{Hartnoll:2009sz,McGreevy:2009xe}. We assume that such materials —considered in an appropriate regime of their defining quantities—  can be described by non-relativistic hydrodynamics~\cite{Hartnoll:2007ih}. In this regime, conserved quantities such as energy, momentum, and electric charge are conserved and lead to long-lived modes, for example sound modes. Outside of this hydrodynamic regime, such materials may contain non-hydrodynamic modes. Some of those may be long-lived as well. The quasinormal mode frequencies as a function of spatial momentum, which we compute in this work, holographically correspond to dispersion relations for the long-lived and short-lived modes which may occur in such materials. However, the system we study in this work is only a step towards non-relativistic descriptions of such materials. This is because we are working in a background which imposes that time scales just like spatial directions. But we will see that our system still singles out the time direction, allows for arbitrarily high velocities, and has further non-relativistic properties discussed below. So we are studying a putative two-dimensional material, sharing symmetries and hydrodynamic description with real materials. We are not attempting to describe the microscopic details of such real world materials.

\section{Ho\v{r}ava Gravity}\label{horSec}
In this section, we will summarize those aspects of Ho\v{r}ava Gravity of relevance to our calculation of QNMs. Our main interest are all the QNMs of fluctuations around a particular analytically known Ho\v{r}ava black brane solution found by Janiszewski~\cite{Janiszewski2015}. 
We close this section with a review of the (numerical) shooting method with which then the Ho\v{r}ava QNMs of the axial sector are computed. In the hydrodynamic limit, axial and polar\footnote{That previous analysis specialized to the case of vanishing coupling $\lambda=0$ for polar QNMs. In this work, we lift this severe restriction and analyze QNMs for nonzero $\lambda$.} transport quantities have been found analytically in \cite{Davison2016}\footnote{In~\cite{Davison2016}, our {\it axial} modes are there called {\it vector} modes, and our {\it polar} modes are there called {\it scalar} modes.}, which will serve as a check of our numerical solutions. 

\subsection{Action, coupling constants \& field content}
We will specialize to a particular low energy solution of classical Ho\v{r}ava Gravity~\cite{Janiszewski2015}. The degrees of freedom are represented by $G_{IJ}$, $N^I$, and $N$, which are constituents of the ADM decomposition of the spacetime metric, $g_{XY}$\footnote{Capitalized roman letter indices $I,J,K,\ldots$ are non-temporal indices (ie not $t$). Capitalized roman letter indices $X,Y,Z$ are bulk spacetime indices.}; where we will be using the mostly positive metric convention\footnote{The mostly positive convention (-,+,+,+).}.
In terms of spacetime coordinates $x^X=\{t,r,x,y\}$ the line element then is given by 
\begin{equation}\label{eqMetricHor}
ds^2=g_{XY}dx^X dx^Y=-N^2 {dt}^2+G_{IJ} ({dx}^I+N^I dt) ({dx}^J+N^J dt) \, .
\end{equation}

In (3+1) dimensions the analytically known metric solution~\cite{Janiszewski2015} satisfies the equations of motion generated by the variation of the following action of Ho\v{r}ava Gravity:  
\begin{equation}\label{eqActionHor}
S_\text{Ho\v{r}ava}=\int  \frac{d^4x\sqrt{|G|}}{16 \pi G_H} \left( K_{IJ} K^{IJ}-(1+\lambda) K^2 + (1+\beta)(R-2 \Lambda)+\alpha \frac{(\nabla_I N) (\nabla^I N)}{N^2} \right)\, , 
\end{equation}
with 
\begin{equation}\label{eqIntrinsicK}
K_{IJ}\equiv \frac{1}{2 N} (\partial_t G_{IJ}-\nabla_I N_J-\nabla_J N_I) \, .
\end{equation}
$K_{IJ}$, $G_{IJ}$, $N^I$ and $N$ are the extrinsic curvature tensor, spatial metric, shift vector, and lapse function, respectively. $K$ is the trace of $K_{IJ}$. $G$ is the determinant of $G_{IJ}$. The lowering and raising of spatial indices are carried out by contraction with $G_{IJ}$ and $G^{IJ}$, respectively, while $\nabla_{I}$ is the covariant derivative defined with respect to the spatial metric $G_{IJ}$. In order for the propagation speeds of the graviton to be strictly positive the coupling constants,$(\alpha,\beta,\lambda)$, must obey the following inequalities \cite{BHATTACHARYYA2013,Griffin2013}:\par 
\begin{equation}\label{eqInequalitiesHorCoupling}
\beta>-1 ~~\text{and}~~ 0\leq \alpha \leq 2 (1+\beta) ~~\text{and}~~ \lambda \geq 0 ~~\text{or}~~ \lambda \leq -\frac{2}{3} \, .
\end{equation}
These {\it allowed} regions of parameter space are represented by white surfaces plus the blue line along the $\beta$-axis in the ($\beta$,$\lambda$)-parameter plots Fig.~\ref{fig:ParameterPolarPlots} and Fig.~\ref{fig:ParameterAxialPlots}; forbidden regions are colored red. 

\subsection{Ho\v{r}ava black brane background solution}

Taking $\alpha=0$ and $\Lambda=-3/L^2$ (as usual, the AdS radius $L$ can be set to unity, $L=1$, by scaling symmetries of the equations of motion), we here review the aforementioned asymptotically ${AdS}_4$ black brane solution to Ho\v{r}ava gravity found in~\cite{Janiszewski2015}. $G_{IJ}$, $N^I$, and $N$  are known analytically and given by\footnote{The radial coordinate~$r$, spatial momentum~$k$, and frequency~$\omega$ carry units of  $length$, $length^{-1}$, and $length^{-1}$ respectively.}
\begin{equation}\label{eqADSHorBlackGroundMetric}
G_{IJ} = \left( \begin{matrix} {\left( \frac{r_h^3}{r(r_h^3-r^3)} \right)}^2 & 0 & 0 \\0 & \frac{1}{r^2} & 0 \\0 & 0 & \frac{1}{r^2}\end{matrix} \right) \qquad N=\frac{1}{r} \left( 1-\frac{r^3}{r_h^3} \right) \qquad N_I=\left(\frac{r \sqrt{1+\beta}}{r_h^3-r^3} \right) \, ,
\end{equation}
which notably is independent from the coupling $\lambda$. 
The AdS radius is $L=1$, the time coordinate is $t\equiv x^0$, spatial coordinates $x\equiv x^2$, $y\equiv x^3$, and the radial coordinate is $r\equiv x^1$. The AdS boundary lies at $r=0$ and the universal horizon at $r=r_h$. This horizon is a trapping surface from which none of the Ho\v{r}ava Gravity fields can escape. In particular, the universal horizon is also the sound horizon\footnote{We define the sound horizon of a field as the location along the radial AdS-coordinate (excluding the AdS-boundary), at which the fluctuation equation for that field contains singular coefficients when the leading derivative is normalized to have coefficient 1. For example, $\phi''(r)+b(r)/(r-r_c)\phi'(r)+ b(r)/(r-r_c)^2\phi(r)=0$ has a sound horizon at $r=r_c$, if $b(r)$ and $c(r)$ can each be expanded in a Taylor series.} for the spin-0 graviton, which travels with infinite speed in this particular solution.
Since Ho\v{r}ava Gravity is non-relativistic, there exists another horizon: the spin-2 sound horizon at $r=r_h / 2^{\frac{1}{3}}$, which is a trapping surface for the spin-2 graviton. There is also the Killing horizon located at $r=r_k$~\cite{Davison2016}.  
The temperature is determined by the universal horizon~\cite{Janiszewski2015} and is given by
\begin{equation}\label{eq:T}
T=\frac{3 \sqrt{1+\beta}}{4\pi r_h } \, .    
\end{equation}

In general, solutions to Ho\v{r}ava Gravity can exhibit Lifshitz scaling symmetry under the scaling with a constant $\kappa$: $t\to \kappa^z t$, \, $x^I\to \kappa x^I, \, r\to \kappa r$ with the dynamical exponent $z$, leading to the asymptotic scaling $ds^2\sim \frac{dt^2}{r^{2z}}+\frac{{dx^I}^2+dr^2}{r^2}$~\cite{Griffin2013}. The metric solution in Eq.~\eqref{eqADSHorBlackGroundMetric} is an example with $z=1$, which means that time and space coordinates scale the same way. However, time and spatial coordinates are still distinct because there exists a time-like vector (or gradient of the khronon) specifying a preferred time-foliation. To see this symmetry explicitly it is helpful to write Ho\v{r}ava Gravity as Einstein-\AE{ther} Theory with a particular constraint (hypersurface orthogonality)~\cite{Davison2016,BHATTACHARYYA2013}, as we will see below in Sec.~\ref{sec:AE}. Hence, the solution~\eqref{eqADSHorBlackGroundMetric}, like any typical solution to Ho\v{r}ava Gravity, is only invariant under those diffeomorphisms preserving the time-foliation.  

\subsection{Ho\v{r}ava black brane perturbations}

One may perturb the metric (\ref{eqMetricHor}) around a background value of (\ref{eqADSHorBlackGroundMetric}) to linear order by a metric perturbation $h_{XY}$ 
\begin{equation}\label{eqPerturbedMetric}
g_{XY}^p = g_{XY}+\epsilon~h_{XY}(t,r,x^I) \, ,
\end{equation}
where $\epsilon \ll 1$. 
Requiring a vanishing variation of the action with respect to
$h_{XY}$ generates 10 coupled  linear equations of motion for the 10 independent components of $h_{XY}$. Since our action~\eqref{eqActionHor} and background metric~\eqref{eqADSHorBlackGroundMetric} have a translational symmetry in $x,y$-directions  (\ref{eqADSHorBlackGroundMetric}), we make the standard plane wave ansatz of Fourier expanding into momentum space modes ${\tilde{h}}_{XY}(r;\omega,\vec{k})$.\footnote{Recall that $x\equiv x^2$ and $y\equiv x^3$.}
Since action and metric also are invariant under spatial rotations in the $x,y$-plane, without loss of generality we choose the momentum vector to point into the $y$-direction
\begin{equation}\label{eqFourierHor}
h_{XY}(t,r,x^I,r) = \int d\omega dk~e^{\boldmath{i} (k y-\omega t)}~\frac{{\tilde{h}}_{XY}(r;\omega,k)}{r^2} \, .
\end{equation}
Once (\ref{eqFourierHor}) is substituted into the equations of motion we find that the 10 equations of motion decouple into a set of 7 equations for metric components which are odd under parity ({\it axial}) and a set of 3 equations for metric components which are even under parity ({\it polar})~\cite{Davison2016}. For the QNMs in the Ho\v{r}ava case we only concern ourselves with the set of 3 equations of motion for the {\it axial} fields ${\tilde{h}}_{xy}$, ${\tilde{h}}_{xt}$, and ${\tilde{h}}_{xr}$, which are odd under parity. After a radial diffeomorphism, ${\tilde{h}}_{xr}(r;\omega,k)$ can be set to vanish. A linear combination of these fields turns out to be a gauge invariant master field, $\psi(r;\omega,k) = (\omega h_{xy}+k h_{tx})$, and obeys the following single equation of motion\footnote{Here $\nu$ and $k$ dependence in $\psi$ is suppressed.} found in \cite{Davison2016}
\begin{multline}\label{eqHorMasterEOM}
\bigg[ q^4 z^2 (-2+z^3)^2 (-1+z^3)+2 {\nu}^2 \big[ -2 {\nu}^2 z^2-i \nu z^4 (-5+z^3)+(-2+z^3)^2 (1+2 z^3) \big] \\ + q^2 \big[{\nu}^2 z^2 (8-8 z^3 +z^6) -2(2-3z^3+z^6)^2+i \nu z^4 (-10+6 z^3 +z^6) \big] \bigg] \psi(z) \\ +z(-2+z^3) \big[ 2 q^2 (-1+z^3) (2+z^3 (-3-i \nu z +z^3)) \\ + {\nu}^2 (4+z^3 (-12-2 i \nu z + z^3)) \big] \psi'(z) \\ z^2 (-2+z^3)^2 (-1+z^3) \big[ {\nu}^2 + q^2 (-1+z^3) \big] \psi''(z) = 0\, ,
\end{multline}
where the following variables have been rescaled to be dimensionless
\begin{equation}\label{eqHorMasterEOMVars}
z \equiv \frac{2^{1/3} r}{r_h} \, ,\qquad 
q \equiv \frac{r_h}{2^{1/3}} k \, ,\qquad \nu \equiv \frac{r_h}{2^{1/3} \sqrt{1+\beta}} \omega \, .
\end{equation}
Eq.~\eqref{eqHorMasterEOM} is the master equation of motion for axial perturbations which we are going to solve in the remainder of this section to extract axial QNMs.

\subsection{Shooting Method}
We intend to find QNMs, that is, we search for those solutions to fluctuation equations which satisfy two conditions: (i) modes are ingoing at the sound horizon, and (ii) vanish at the boundary, i.e. satisfy a Dirichlet boundary condition. With the rescaled-$z$ coordinate~\eqref{eqHorMasterEOMVars}, the sound horizon is located at $z=1$, where the equation of motion~\eqref{eqHorMasterEOM} has a regular singular point; in fact, this regular singular point is the reason for us to call this location a sound horizon. We make the following near sound horizon ansatz
\begin{equation}\label{eqPsiNearHorExpansion}
\psi(z;\nu,q) \equiv (1-z)^\alpha F(z;\nu,q) = (1-z)^\alpha \sum_{n=0}^{\infty} f_n (\nu,q) (1-z)^n \, ,
\end{equation}
separating the regular part $F$ and irregular part of $\psi(z;\nu,q)$ from each other. Solving the equation (\ref{eqHorMasterEOM}) with the ansatz (\ref{eqPsiNearHorExpansion}), one finds two $\alpha$'s that satisfy Eq.~(\ref{eqHorMasterEOM}) at the first non-zero order in the near-horizon expansion: $\alpha=-1$ or $\alpha=-\frac{2 i \nu}{3}$. We choose the latter which corresponds to the ingoing mode. At the other expansion orders one can recursively solve for the $f_n (\nu,q)$ coefficients. As usual, the series in Eq.~(\ref{eqPsiNearHorExpansion}) is asymptotic, though one can assume validity for a small region around the sound horizon. Since there are singularities at $z=1$ and $z=0$ we numerically solve Eq~(\ref{eqHorMasterEOM}) with Mathematica's NDSolve function on the restricted computational domain of\footnote{z=1 is the location of the sound horizon.} $z \in [{dr}_b,1-{dr}_h]$ where ${dr}_b,{dr}_h \ll 1$. It as been shown that for exceedingly small values of ${dr}_h$ the (\ref{eqPsiNearHorExpansion}) ansatz fluctuates rapidly and can create large numerical errors~\cite{Kaminski:2009ce}, which guides our choice of $dr_h$ here. The boundary conditions, $\psi(1-{dr}_h)$ and $\psi'(1-{dr}_h)$, are then provided by the horizon expansion~(\ref{eqPsiNearHorExpansion}). For an arbitrary value of $\nu$ and $q$, this solution is not a quasinormal mode, {i.e.}~$\psi({dr}_b) \neq 0$. We use Mathematica's FindRoot to find $\nu$ such that $\psi({dr}_b;q) = 0$. When compared to the pseudospectral method's modes (Fig. \ref{fig:AxialCPPlot} and Fig. \ref{fig:PolarCPPlot}), the shooting method used here was found to be numerically less stable, especially for QNMs with larger momentum. Hence, in the next section we will switch to pseudospectral methods. 

\subsection{Ho\v{r}ava Gravity axial QNMs}\label{sec:horavaAxialQNMs}
It turns out that the axial QNMs $\nu$ found with the shooting method are numerically equal the QNMs one would find for an asymptotically AdS$_4$ Schwarzschild black brane within Einstein Gravity, {i.e.} $\nu^{\text{Ho\v{r}ava}}_{\text{axial QNM}} = \nu^{\text{Einstein}}_{\text{axial QNM}}$. However, the  QNM frequencies $\nu$ are scaled by a factor of $\sqrt{1+\beta}$ and $\beta$ vanishes in GR. So, up to numerical errors we empirically find the relation
\begin{equation}
    \omega^{\text{Ho\v{r}ava}}_{\text{axial QNM}} = \sqrt{1+\beta}\, \omega^{\text{Einstein}}_{\text{axial QNM}} \, .
\end{equation}
This particular $\beta$ factor is in  fact the speed of the spin-2 graviton~\cite{Davison2016}. At $\beta=0$ the spin-2 graviton travels at unit speed, and with respect to just the axial perturbations, the theory returns to being relativistic. At small momentum our lowest lying mode agrees with the diffusion mode found in a hydrodynamic approximation given in Eq.~(3.31) through (3.34) of~\cite{Davison2016}.

It must be mentioned that we attempted to find polar QNMs, however Mathematica's NDSolve used in the shooting method failed to find a converging solution to the fluctuation equations of motion.\footnote{This is due to an apparent pole at $r=\frac{r_h}{2^{1/3}}=r_s$. This still happened despite the polar mode equations of motion not having any factors of $(r-r_s)$ in the perturbation coefficients, which would have indicated regular/irregular poles at $r_s$.} 
It is difficult to determine indicial exponents in general for this coupled system, and to then separate the singular from the regular part of the fluctuations. 
To circumvent these problems, we decided to calculate axial and polar QNMs in an equivalent theory, 
Einstein-\AE ther Gravity, using a different technique, namely pseudospectral methods. The equivalence of these two theories holds under the constraint of hypersurface orthogonality, discussed in Sec.~\ref{sec:AE}. Hence, QNMs found in the two theories are expected to be identical\footnote{Only a smaller subset of the frequencies found with spectral methods were also found via the shooting method.}. A comparison between axial QNMs computed in both theories is displayed in table~\ref{tab:axialQNMComparison}. For the shooting 9 orders in the horizon expansion have been taken into account, the horizon and boundary cutoffs were chosen as $r=1-10^{-3}$ and $r=10^{-3}$, respectively. For the computation of the Einstein-{\AE}ther QNMs a grid of size $N_{grid}=80$ was compared to one with $N_{grid}=100$ in order to determine convergent quasinormal modes, as described in appendix~\ref{sec:convergence}. The percentage of deviation $d$ of the Ho\v{r}ava Gravity QNM frequencies from the Einstein-{\AE}ther Theory QNMs, for that calculation see Sec.~\ref{sec:AE}, is given by 
\begin{equation}\label{eq:d}
    d= 2\frac{\nu_H-\nu_{\AE}}{\nu_H+\nu_{\AE}} \times 100\% \, . 
\end{equation}
Up to numerical errors which are at worst on the order of $10^{-6}$ percent, we find agreement between QNMs of {\it hypersurface-orthogonal} Einstein-\AE ther Theory and Ho\v{r}ava Gravity in the five axial QNMs we have checked for many values of momentum $q$, and report only examples in table~\ref{tab:axialQNMComparison}. This serves as a check of the equality of the QNM spectra 
\begin{equation}\label{eq:QNMEquality}
\nu_{\text{QNM}}^{\text{Ho\v{r}ava}} = \nu_{\text{QNM}}^{\text{hypersurface-orthogonal Einstein-\AE ther}}\, ,
\end{equation}
which will be discussed in the next section.
\begin{table}[]\footnotesize
\centering
\begin{tabular}{|c|c|c|c|}
\hline 
 q & Ho\v{r}ava $\nu$ & 
 deviation $d$ from Einstein-\AE ther [in \%]\\
 \hline \hline
 0.1 
 & $2.040790625396911\times10^{-12} - 0.003336148565969461 i$ 
 & $5.92777\times10^{-7} + 8.05655\times10^{-8} i$ 
 \\
 \hline
 0.1 
 & $1.850804328436749 - 2.6634620817084143 i$ 
 & $-1.04789\times10^{-7} + 9.46626\times10^{-8} i$ 
 \\
 \hline
 0.5 
 & $6.311321466437316\times10^{-13} - 0.08515846254765713 i$ 
 & $2.56115\times10^{-10} + 7.41127\times10^{-10} i$ 
 \\
 \hline
 0.5 
 & $1.8825874031004601 - 2.654086452911871 i$ 
 & $-1.04405\times10^{-7} + 6.75017\times10^{-8} i$ 
 \\
 \hline
 1.0 
 & $5.842004644938402\times10^{-13} - 0.3665129458025562 i$ 
 & $3.33525\times10^{-10} + 1.59394\times10^{-10} i$ 
 \\
 \hline
\end{tabular}
\caption{Sample comparison of axial QNM frequencies $\nu$ computed from Ho\v{r}ava Gravity using the shooting method and QNMs computed from Einstein-{\AE}ther Theory using spectral methods from Sec.~\ref{sec:AE}. The deviation $d$ is defined in Eq.~\eqref{eq:d}, the samples are the hydrodynamic mode at momenta $q=0.1,\,0.5,\, 1.0$, and the lowest non-hydrodynamic mode at $q=0.1,\, 0.5$.}
\label{tab:axialQNMComparison}
\end{table}
%

\section{Einstein-\AE ther Theory}\label{sec:AE}
Einstein-\AE ther Theory differs from General Relativity by the addition of a scalar matter field $\phi$, which is termed \AE ther field. We need a particular Einstein-\AE ther Theory, namely one that is equivalent to Ho\v{r}ava gravity. This requires the \AE ther field to define a timelike vector field which is hypersurface orthogonal. That goal is achieved by the following definition of the timelike vector 
\begin{equation}\label{eqKhrononField}
u_\mu=\frac{{\partial}_\mu \phi}{\sqrt{-({\partial}_\nu \phi) ({\partial}^\nu \phi)}},
\end{equation}
which is a timelike 
hypersurface orthonormal unit vector matter field~\cite{Jacobson2000,Bhattacharyya:2008jc}. This vector field also breaks diffeomorphism invariance down to foliation preserving diffeomorphisms. 
In this context, the scalar $\phi$ is often referred to as {\it khronon}, determining the time foliation. 
Lower case Greek indices $\mu,\nu$ denote spacetime indices. As coordinates, we choose $x^\mu$ with $x^0=v,~x^1=r,~x^2=x,~x^3=y$, and use the mostly positive metric convention, (-,+,+,+). 

A (3+1)-dimensional Einstein-\AE ther action 
has been constructed~\cite{Jacobson2000}, which includes four quadratic derivative terms of $u_\mu$,\footnote{We omit here the constraint term, $\lambda_{\text{\AE}}(u^2+1)$, since Eq.~\eqref{eqKhrononField} incorporates this unit constraint by construction.}
\begin{multline}\label{eqAetherAction}
S_{\text{\AE ther}}=\frac{1}{4 \pi G_{ae}} \int d^4x  \sqrt{-g} \bigg( R-2 \Lambda+c_4 (u^\mu {\nabla}_\mu u^\nu)(u^\sigma {\nabla}_\sigma u_\nu)-c_3 ({\nabla}_\mu u^\nu)({\nabla}_\nu u^\mu) \\ -c_2 ({\nabla}_\mu u^\mu)^2-c_1 ({\nabla}^\mu u^\nu)({\nabla}_\mu u_\nu) \bigg) \, ,
\end{multline}
where $g$ is the determinant of the metric $g_{\mu \nu}$, and $R$ is the Ricci Scalar of the metric $g_{\mu \nu}$. $c_1$ is redundant once we constrain $u$ to be hypersurface orthogonal by construction. This allows us to rearrange (\ref{eqAetherAction}) whose coupling constants then appear as linear combinations of $c_1$, $c_2$, $c_3$, and $c_4$ as $c_{13}=c_1+c_3$, $c_2$, and $c_{14}=c_1+c_4$ \cite{BHATTACHARYYA2013}, and we set $c_1=0$ without loss of generality. This equates Einstein-\AE ther theory to low energy Ho\v{r}ava Gravity after the identification $-N=\delta^\mu_0 u_\mu$, and with the following coupling constants for the respective theories \cite{Barausse2011}
\begin{equation}\label{eqCouplingConstantsRelations}
\frac{G_H}{G_{ae}}=1+\beta=\frac{1}{1-c_3}\, ,\qquad 1+\lambda=\frac{1+c_2}{1-c_3}\, ,\qquad \alpha=\frac{c_4}{1-c_3} \, .
\end{equation}
Analogously to our perturbative treatment in the previous section around Eq.~(\ref{eqMetricHor}), also here we will investigate linear perturbations around the black brane solution with $\alpha=0$ and $\Lambda =-3$ \cite{Janiszewski2015}. Our coordinates are similar to Eddington-Finkelstein coordinates~\cite{Janiszewski2015}, as seen in the metric 
\begin{align}\label{eqAetherMatrix}
g_{\mu \nu} & = \left( \begin{matrix} -e(r) & \pm f(r) & 0 & 0 \\ \pm f(r) & 0 & 0 & 0 \\0 & 0 & \frac{1}{r^2} & 0 \\0 & 0 & 0 & \frac{1}{r^2} \end{matrix} \right) \, ,\\ 
u_\mu & =\left(\frac{a^2(r) e(r)+(f(r))^2}{2 a(r) f(r)},\pm a(r),0,0 \right)\, , \nonumber \\ 
e(r) & =\frac{1}{r^2}-\frac{2 r}{r^3_h}-\frac{c_3 r^4}{(1-c_3)r^6_h}\, ,
\qquad f(r)=\frac{1}{r^2}\, ,\qquad a(r) =\frac{r^3_h}{r^3_h r + \left( \frac{1}{\sqrt{1-c_3}}-1 \right) r^4}\, ,\nonumber
\end{align} 
where $r_h$ is the radius of the universal horizon. The the sign choice on $f(r)$ and in $u_\mu$ corresponds to the choice of infalling (lower signs) or outgoing (upper signs) Eddington-Finkelstein-like coordinates~\cite{Janiszewski2015}. For this paper we choose the negative signs ($f(r)=-1/r^2$ and $u_r=-a(r)$). This leads to infalling modes which are regular at their respective sound horizons. 
Note that Eq.~\eqref{eqAetherMatrix} reduces to the Schwarzschild $AdS_4$ metric with the Schwarzschild horizon located at $r=r_{\text{Schwarzschild}}=r_h/2^{1/3}$ when all remaining \AE ther couplings are set to zero, $c_3=0=c_2$, and we choose the vector field to vanish $u_\mu =0$.  
As a side note, remarkably, one can express $\phi$ explicitly by integrating Eq.~\eqref{eqKhrononField} to obtain
\begin{equation}\label{eqKhrononPhiField}
\phi(r,\nu)=\Bigg( \int_{ }^r \frac{2 a(\rho)^2 f(\rho)}{a(\rho)^2 e(\rho)+f(\rho)^2} \, d\rho \Bigg) -\nu \, .
\end{equation}
%

\subsection{Einstein-\AE ther black brane perturbations}

Similar to how we perturbed the Ho\v{r}ava black brane (\ref{eqPerturbedMetric}) we perturb the metric (\ref{eqAetherMatrix}) by adding a ``small" linear term where $\epsilon \ll 1$, we choose
\begin{align}\label{eqPerturbedMetric1}
g_{\mu \nu}^p & = g_{\mu \nu}+\epsilon~h_{\mu \nu}(x^\sigma) \\ \phi^p & = \phi+ \epsilon~ \chi (x^\sigma) \, , \nonumber
\end{align}
where $\chi(x^\sigma)$ is a scalar field. Since $\phi^p$ is still a scalar field after the perturbation, replacing $\phi\to\phi^p$ in Eq.\eqref{eqKhrononField} ensures hypersurface orthogonality and normalization of the now perturbed $u$ vector. The $h_{\mu \nu}$ fields and $\chi$ obey eleven coupled linear equations. A Fourier transformation similar to (\ref{eqFourierHor}), is applied to (\ref{eqPerturbedMetric1}), yielding 
\begin{align}\label{eqFourierAether}
h_{\mu \nu}(x^\sigma) & = \frac{2^{2/3}\sqrt{1+\beta}}{r_h^2} \int d\nu dq~e^{\boldmath{i} \frac{2^{1/3}}{r_h} (q y-\sqrt{1+\beta}~\nu v)}~{\tilde{h}}_{\mu \nu}(r;\nu,q) \, , \\ \chi(x^\sigma) & = \frac{2^{2/3}\sqrt{1+\beta}}{r_h^2} \int d\nu dq~e^{\boldmath{i} \frac{2^{1/3}}{r_h} (q y-\sqrt{1+\beta}~\nu v)}~{\tilde{\chi}}(r;\nu,q) \, . \nonumber
\end{align}

Using (\ref{eqFourierAether}) in the eleven equations of motion, they decouple to two sets of three equations of motion and eight equations of motion which depend on fields
($\tilde{h}_{xt}$, $ \tilde{h}_{xr}$,  $\tilde{h}_{xy}$), and  ($\tilde{h}_{xx}$, $\tilde{h}_{yy}$, $\tilde{h}_{yt}$, $\tilde{h}_{yr}$, $\tilde{h}_{rr}$, $\tilde{h}_{tt}, \tilde{h}_{rt}$, $\tilde{\chi}$), respectively. 
The reason for this decoupling is that the fields ($\tilde{h}_{xt}$, $ \tilde{h}_{xr}$,  $\tilde{h}_{xy}$) are odd under parity, hence we refer to them as {\it axial}. However, the remaining eight fields are even under the parity transformation $x\to-x$, and we refer to them as {\it polar}. Since the perturbation equations are linear in perturbations, they can not couple fields of different symmetry properties~\cite{Miranda2008,Kovtun:2005ev}. 
The coupled fluctuation equations are lengthy, hence we include them in ancillary Mathematica~\cite{Mathematica10} files with this submission.

In the rest of this section, we obtain all QNM results using pseudospectral methods. This method turns out to be more efficient in finding QNMs compared to the shooting method described in Sec.~\ref{horSec}. We apply the general techniques described well in~\cite{Boyd2000}. 
More specifically, the recent Mathematica package for finding AdS quasinormal modes~\cite{Jansen2017} has been very useful for generating and checking our code. 

Pertaining to the polar modes, there are eleven coupled equations of motion, while there are three for axial modes. We convert each set of equations to a linear algebra statement using a Gauss-Lobatto grid, with 80 to 100 grid points. More specifically, the linear algebra problem is a generalized eigenvalue problem where the complex eigenvalues are the quasinormal mode frequencies we seek to find~\cite{Jansen2017}. 
A spectral matrix is constructed as a representation of the problem, and Mathematica's $\mathsf{Eigenvalues[\dots]}$ is used to find the generalized eigenvalues, i.e. the QNMs. We note here that the determinant of the relevant matrix vanishes in general, which obstructs inverting that matrix. Hence the treatment as a generalized eigenvalue problem. 
The procedure for finding convergent quasinormal modes is outlined in appendix~\ref{sec:convergence}.

In order to determine which horizon is relevant to each sector, we determine the regular singular points of the linear system of differential equations. A perturbative analysis of the coefficients in these equations reveals that the coefficients become singular at a certain radial coordinate value, which is the sound horizon relevant for this sector. 
Our domain of integration stretches from the AdS-boundary $r=0$ to the relevant sound horizon, which we set to $r=1$. For the axial modes, this is achieved by fixing $r_h= 2^{1/3}$, because the relevant horizon for axial modes is the spin-2 sound horizon at $r=r_h/2^{1/3}$. Note, that this fixes the temperature, Eq.~\eqref{eq:T} to $T=3/(4\pi 2^{1/3}\sqrt{1-c_3})=3\sqrt{1+\beta}/(4\pi 2^{1/3})$. For the polar modes, the horizon is set to $r=1$ by choosing $r_h=1$, because the relevant horizon for polar modes is the universal horizon, which is also the sound horizon for the spin-0 graviton, $r=r_h$. This fixes the temperature $T=3/(4\pi\sqrt{1-c_3})=3\sqrt{1+\beta}/(4\pi)$ for the polar modes, which is distinct from the axial case temperature by a factor of $2^{1/3}$. Since the temperature is the only scale we fix here (except for the AdS-radius $L$ fixed using scale symmetries of the equations of motion), our QNM results are still general.\footnote{We have checked this explicitly by redefining the radial coordinate, showing that $r_h$ disappears from the equations of motion. Hence the QNM frequencies we report are independent of this choice of horizon location in the units we are using.} 
With these definitions, all our frequencies will be expressed collectively in units of temperature, as nicely realized by the frequency definition Eq.~\eqref{eqHorMasterEOMVars}:
\begin{equation}
    \nu = \left( 
    \frac{3}{4\pi\,2^{1/3}}
    \right )\frac{\omega}{T},
\end{equation}
for axial and polar sector. 

\subsection{Quasinormal mode results} \label{secQNMPseudoSpectral}
We find two sets of QNMs coming from the axial and polar sector, respectively, corresponding to the decoupled equations of motion found in the previous section. Plotted in the complex frequency plane, these QNM frequencies are symmetric about the imaginary axis and have negative imaginary values, which shows perturbative stability of the background (in the large parameter region of couplings, $\lambda,\, \beta$, and momenta $k$, which we computed). 
In Einstein Gravity the speeds and horizons of all polar and axial excitations are identical.
However, here in Einstein-\AE ther or equivalently Ho\v{r}ava Gravity, the excitations travel at distinct speeds and hence ``see'' distinct horizons.  
Axial QNMs are characterized by speed,$\sqrt{1+\beta}$, which is the speed of the spin-2 graviton~\cite{Jacobson2004,Janiszewski2015} and its sound horizon is located at $r_s=\frac{r_h}{2^{1/3}}$~\cite{Janiszewski2015}. The polar sector contains the spin-0 graviton (or khronon) and the remaining components of the metric, all traveling at infinitely large speed as $\alpha \rightarrow0$~\cite{Jacobson2004,Griffin2013,Janiszewski2015}, and the corresponding horizon is $r_h$, the universal horizon~\cite{Janiszewski2015}. 
In Einstein Gravity, the horizon radius of the black brane solution naturally sets a scale. However, with various horizons, we have a choice  where to apply boundary conditions, or to which horizon we normalize other scales, such as our QNM frequencies. As stated above, the computational domain for the axial modes reaches from the boundary $r=0$ to the spin-2 sound horizon $r_s$. For polar modes, the relevant horizon is the spin-0 sound horizon $r_h$. 

\subsubsection{Axial modes}
\begin{figure}[!htb]
 \includegraphics[width=\textwidth]{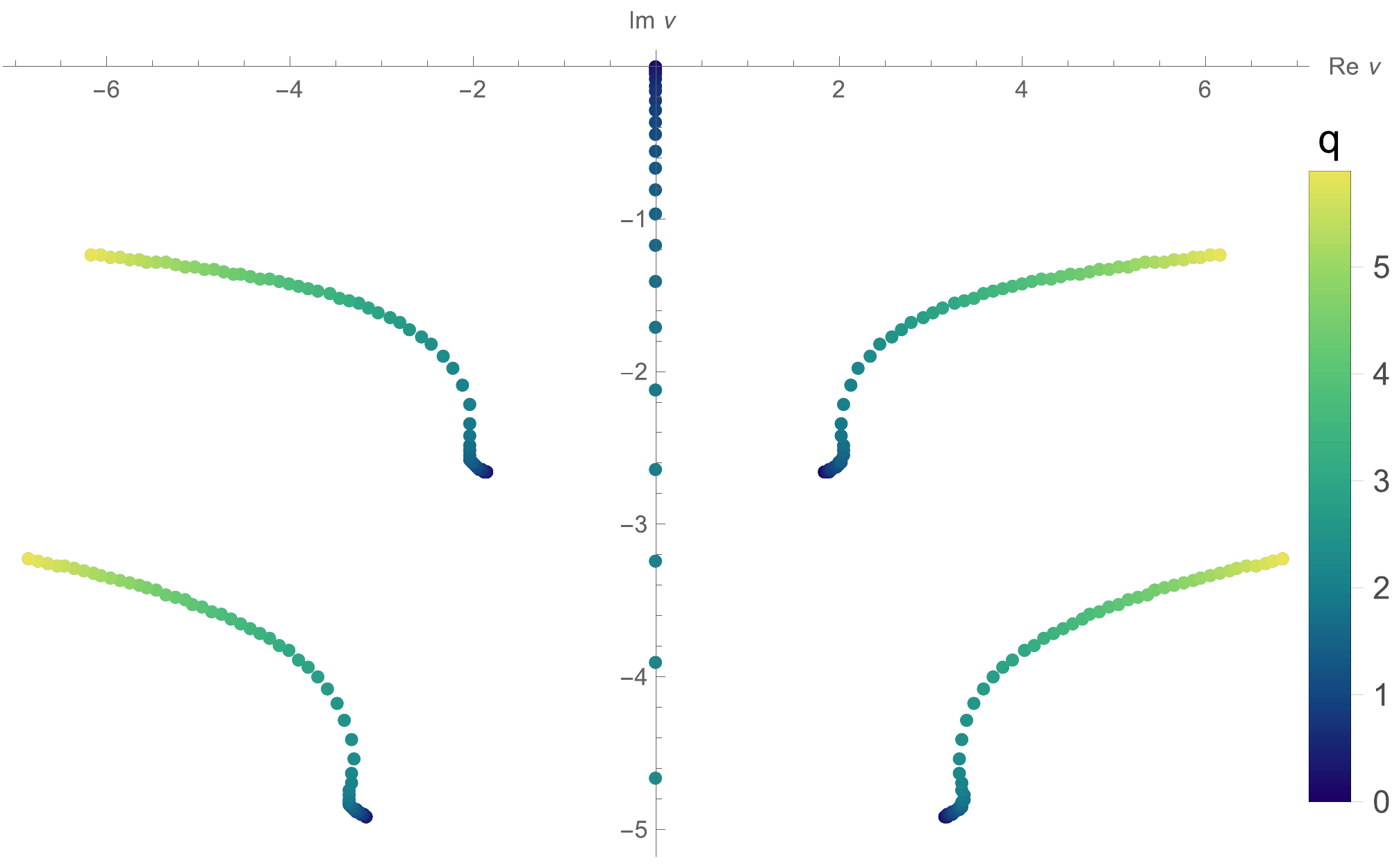}
 \caption{\label{fig:AxialCPPlot}
 Axial modes of Einstein-\AE ther theory: Dimensionless QNM frequencies $\nu$ displayed in the complex frequency plane. 
 Each point corresponds to a dimensionless momentum in the range $0\leq q\leq 6$. 
 }
\end{figure}
For axial modes in this section the relevant causally connected radial domain stretches from the boundary $r=0$ to the spin-2 sound horizon $r_s=r_h/2^{1/3}$. We choose to work at a fixed temperature by $r=r_h/2^{1/3}=1$.  
In Fig.~\ref{fig:AxialCPPlot}, we show axial QNMs corresponding to the axial perturbations $\tilde{h}_{xt}$, $ \tilde{h}_{xr}$, and $\tilde{h}_{xy}$.
The color indicates the magnitude of the dimensionless momentum $q$ in the range $0\leq q\leq 6$. For all axial QNMs, we find that their values measured in the dimensionless frequency $\nu$ agree numerically with the values of QNMs of a Schwarzschild black brane in Einstein Gravity~\cite{Morgan:2009pn,Miranda2008}. This empirical evidence implies that the dimensionful frequencies are related as follows
\begin{equation}\label{eq:axialQNMEquality}
\omega^{\text{axial}}_{\text{Ho\v{r}ava}} = \sqrt{1+\beta}~\omega^{\text{axial}}_{\text{Einstein}}\, ,
\end{equation}
because $\nu=r_h \omega/(2^{1/3}\sqrt{1+\beta})$, where $\beta=0$ for Einstein Gravity. Therefore, also the dispersion relations $\omega(k)$ of this theory are related to those of Einstein Gravity by Eq.~\eqref{eq:axialQNMEquality}.

\begin{figure}[!htb]
\begin{center}
 \includegraphics[width=0.75\textwidth]{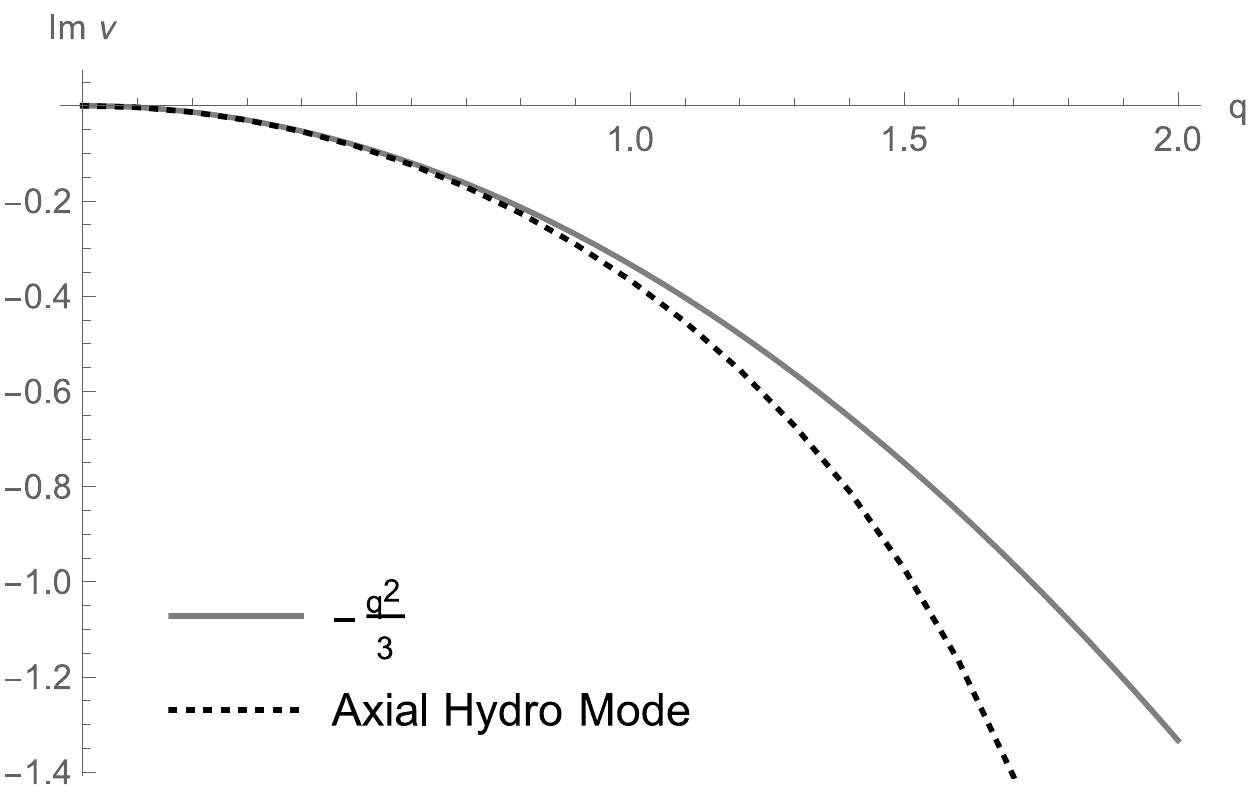}
 \caption{
 \label{fig:AxialHydroPlot}
 Dispersion of hydrodynamic axial mode: Imaginary part of the (dimensionless) frequency $\nu$ associated with the axial hydrodynamic mode as a function of  dimensionless momentum $q$. The numerical result is shown as a dashed line, the hydrodynamic approximation is shown as a solid line. 
 The real part of the mode vanishes.}
\end{center}
\end{figure}

Two distinct types of QNMs appear: hydrodynamic and non-hydrodynamic ones, $\nu_{h}$ and $\nu_{nh}$, respectively. The hydrodynamic QNM obeys the defining relation that its frequency vanishes as the momentum vanishes
\begin{equation}\label{eqHydroCondition}
\lim_{q\to 0^{+}} \nu_h(q) = 0 \, . 
\end{equation}  
Large momentum for this hydrodynamic momentum diffusion mode leads to increasing imaginary frequency indicating increasing dissipation, as expected. 
For the non-hydrodynamic modes, large momenta are leading to large real parts of the frequencies. 
All these axial QNMs are identical to the Ho\v{r}ava QNMs of section~\ref{horSec} up to numerical errors, as indicated by the examples in table~\ref{tab:axialQNMComparison}. 
The hydrodynamic QNM frequency has vanishing real part and its imaginary part monotonically increases with momentum. At sufficiently small momentum $q<1$, our numerically computed frequencies agree well with the analytically predicted momentum diffusion~\cite{Davison2016}
\begin{equation}\label{eq:VectorHydroDispersion}
\nu_{h}(q) = -{i} D q^2 + \mathcal{O}(q^4) \, ,
\end{equation}
up to corrections of order $\mathcal{O}(q^3)$, and 
with the diffusion coefficient $D=1/3$. That value is consistent with the relativistic value $D=\eta/(\epsilon+P)$~\cite{Herzog:2003ke}, which for the Ho\v{r}ava black brane, Eq.~\eqref{eqAetherMatrix}, evaluates to $\frac{1}{3}(1+\beta)^{-1/2} r_h/2^{1/3}$~\cite{Davison2016}. 
The relation~\eqref{eq:VectorHydroDispersion} has been verified numerically, and is also visualized in the fit shown in Fig.~\ref{fig:AxialHydroPlot}. That figure shows the imaginary part of the hydrodynamic mode
\begin{figure}[!htb]
 \includegraphics[width=\textwidth]{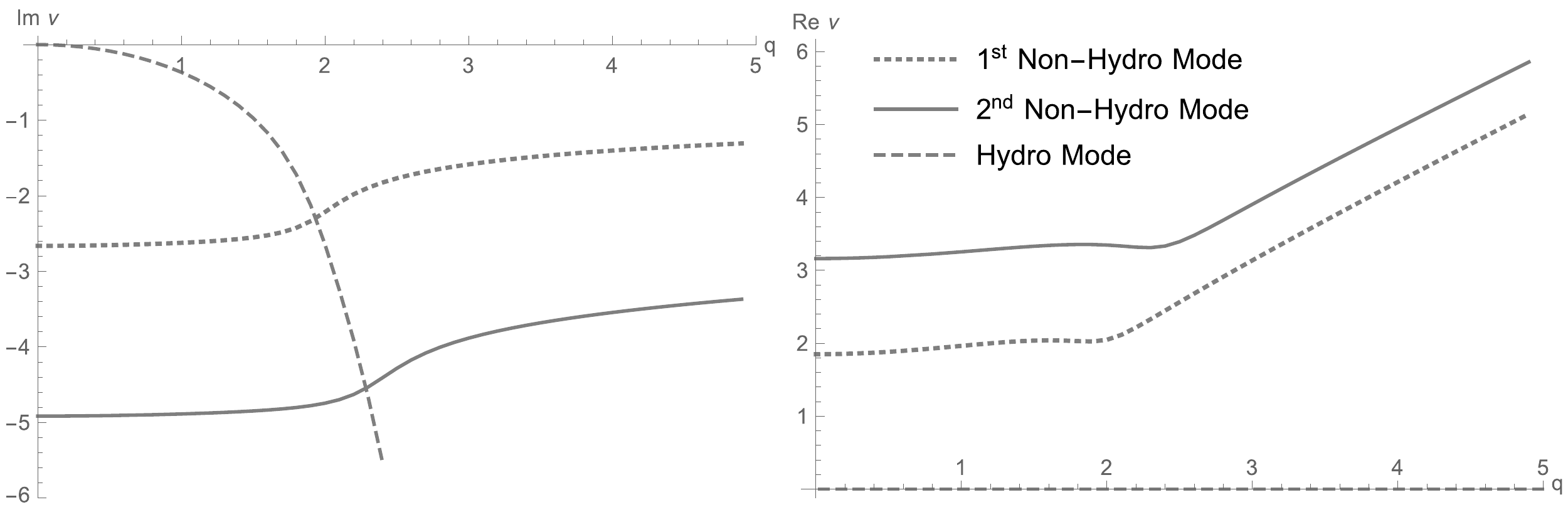}
 \caption{
 \label{fig:AxialNonHydroPlot}
 Dispersion of non-hydrodynamic axial modes compared to hydrodynamic axial mode:
 Imaginary part (left plot) and real part (right plot) of the (dimensionless) frequency $\nu$ associated with the three lowest-lying axial quasinormal modes as a function of  dimensionless momentum $q$. 
 }
\end{figure}
Fig.~\ref{fig:AxialNonHydroPlot} shows a comparison of the dispersions of the two lowest non-hydrodynamic modes with the hydrodynamic diffusion mode. It is interesting to note that around a momentum of $q\approx 2$ the diffusion mode has an imaginary part rapidly growing with momentum, indicating that diffusion modes with larger momentum are rapidly damped. The non-hydrodynamic modes on the other hand display monotonically decreasing imaginary part with increasing momentum. This leads to a crossing between the lowest non-hydrodynamic mode and the diffusion mode at $q_\text{cross}\approx 1.9$; this occurs at $\operatorname{Im}(\nu_{h}(q_\text{cross}))=\operatorname{Im}(\nu_{nh}(q_\text{cross}))$. While the late time behavior of the system for momenta $q<1.9$ is governed by the diffusion mode, the lowest non-hydrodynamic mode dominates the late time behavior for excitations with $q<1.9$. The real part of the non-hydrodynamic quasinormal modes grows linearly for momenta outside the hydrodynamic regime, i.e. with $q>2$. All these observations mirror the behavior of relativistic dispersion relations extracted from holography by virtue of the empirical relation Eq.~\eqref{eq:axialQNMEquality}. In table~\ref{tab:smallQAxial} we collect the expansion coefficients parametrizing the dispersion relations of the five lowest quasinormal modes (and for their mirror images over the imaginary frequency axis). We allow the expansion coefficients to be complex valued. The hydrodynamic diffusion mode has only imaginary coefficients in agreement with the requirement that the corresponding transport coefficient, the diffusion constant $D$, be real valued.
\begin{table}[]
\centering
\begin{tabular}{|c|c|c|c|c|}
\hline 
 $\nu _0$ & $\nu _1$ & $\nu _2$ & $\nu _3$ & $\nu _4$ \\
 \hline \hline
 $0.$ & $0.$ & $0.  -0.333333 i$ & $0.$ & $0.  -0.028111 i$ \\
 \hline
 $\pm1.84942  -2.66385 i$ & $0.$ & $\pm0.138538  +0.039062 i$ & $0.$ & $\mp0.023605-0.000358 i$ \\
 \hline
 $\pm3.16126  -4.91642 i$ & $0.$ & $\pm0.105299  +0.027848 i$ & $0.$ & $\mp0.01297-0.000165 i$ \\
 \hline
 $\pm4.46435  -7.16754 i$ & $0.$ & $\pm0.08816  +0.023133 i$ & $0.$ & $\mp0.008838-0.000102 i$ \\
 \hline
 $\pm 5.76525  -9.41808 i$ & $0.$ & $\pm 0.077319  +0.020279 i$ & $0.$ & $\mp0.006816-0.000059 i$ \\
 \hline
\end{tabular}
\caption{\label{tab:smallQAxial}
Shown here are the expansion coefficients for small momentum axial modes, where $\nu(q) \approx \sum_{m=0}^{4} q^m \nu_m$.}
\end{table}

\subsubsection{Polar modes}
\begin{figure}[]
 \includegraphics[width=\textwidth]{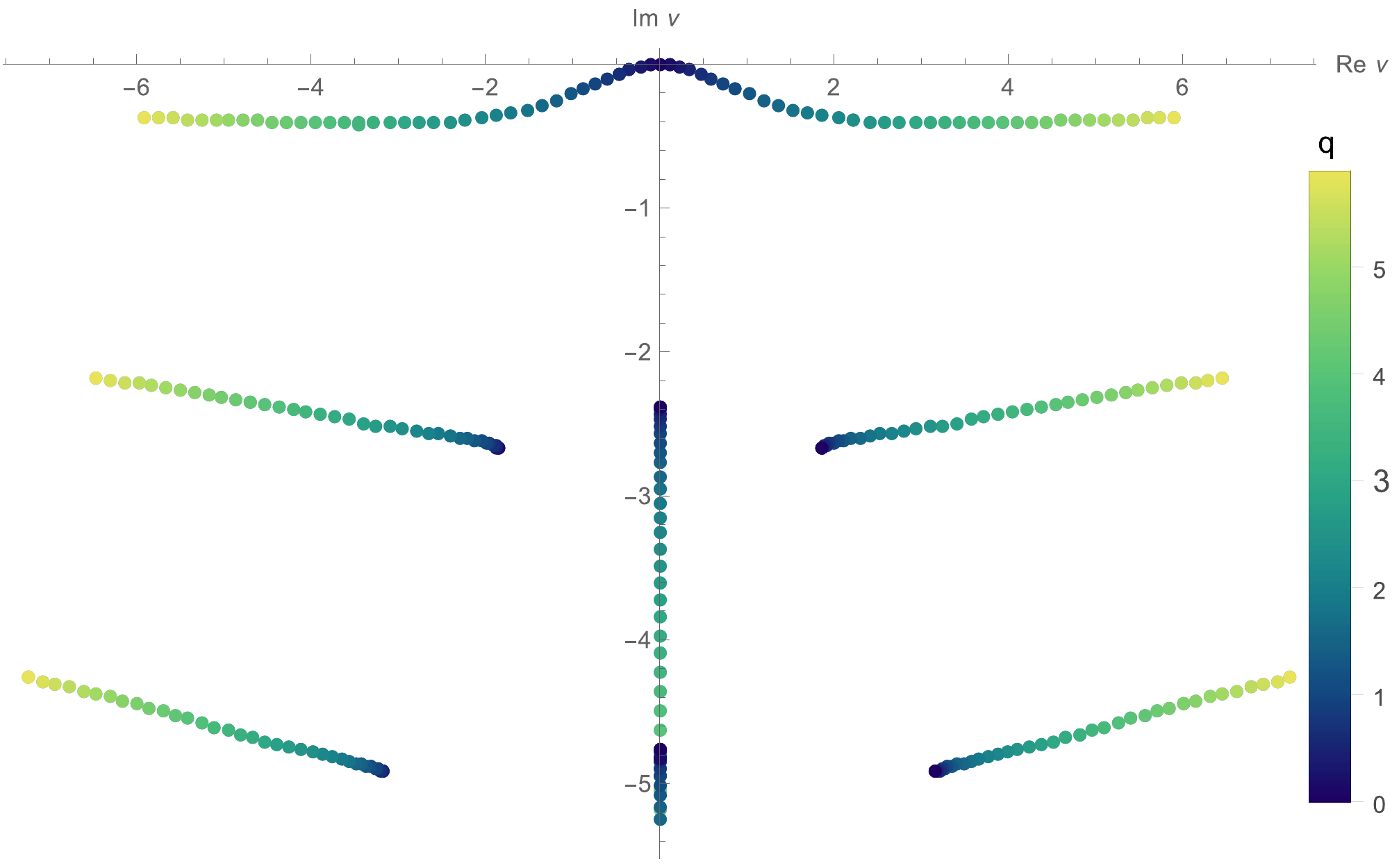}
 \caption{\label{fig:PolarCPPlot}
 Polar modes of Einstein-\AE ther theory: Dimensionless QNM frequencies $\nu$ displayed in the complex frequency plane. 
 Each point corresponds to a dimensionless momentum in the range $0\leq q\leq 6$. The khronon modes move along the imaginary frequency axis with increasing momentum.}
\end{figure}
Polar QNMs correspond to the perturbations $\tilde{h}_{xx}$, $\tilde{h}_{yy}$, $\tilde{h}_{yt}$, $\tilde{h}_{yr}$, $\tilde{h}_{rr}$, $\tilde{h}_{tt}, \tilde{h}_{rt}$, and $\tilde{\chi}$. 
The eight~\footnote{Counting includes mirror QNMs with opposite sign of the real part of the frequency.} lowest polar QNMs are displayed in Fig.~\ref{fig:PolarCPPlot}. 
Similar to the axial QNMs, the polar QNMs can be scaled and expressed in dimensionless units (\ref{eqHorMasterEOMVars}). As indicated in Fig.~\ref{fig:ParameterPolarPlots}, the QNM spectrum looks different for the two cases $\lambda=0$ and $\lambda\neq 0$. With $\lambda = 0$, the polar QNMs are equivalent to black brane QNMs, up to a factor of $\sqrt{1+\beta}$, which can be seen by comparison to the Einstein Gravity $AdS_4$ QNMs computed in~\cite{Morgan:2009pn,Miranda2008}. Just like the axial QNMs, also the polar QNMs numerically agree with the Einstein Gravity QNMs of an $AdS_4$ Schwarzschild black brane when measured in the dimensionless frequency $\nu$, so
\begin{equation}\label{eq:polarQNMEquality}
\omega^{\text{polar}}_{\text{Einstein-\AE ther}} = \sqrt{1+\beta}~\omega^{\text{polar}}_{\text{Einstein}}\, \quad (\lambda=0) .
\end{equation}

When $\lambda \neq 0$, additional purely dissipative non-hydrodynamic modes are found along the imaginary frequency axis. We refer to these as {\it khronon modes}, because they are associated with the fluctuations of the scalar field, $\chi$. This can be confirmed by artificially setting the metric fluctuations to zero and the remaining frequencies found with the spectral method are indeed the same non-hydrodynamic frequencies found when $\lambda \neq 0$ and $h_{\mu \nu} = 0$. Remarkably, the location of the khronon modes and the QNMs associated with the metric appear to be independent of each other, independent of the value of the scalar coupling $\lambda$. This implies that Eq.~\eqref{eq:polarQNMEquality} is true for the QNMs associated with metric fluctuations even at $\lambda\neq 0$. 

Similar to the axial QNMs, there are both hydrodynamic and non-hydrodynamic modes present in the polar sector. 
There are two polar hydrodynamic QNMs, $\nu_{hs}(q)$, which obey the following dispersion relation up to numerical errors: 
\begin{equation}
\label{eqScalarHydroDispersion}
\lim_{q\to 0^{+}}\nu_{hs}(q)= \pm v_s q -i \Gamma q^2 =\pm \frac{1}{\sqrt{2}} q-i \frac{1}{6} q^2 +\mathcal{O}(q^3) \, ,
\end{equation}
which can be rewritten in terms of physical frequency and momentum
\begin{equation}
\label{eqScalarHydroDispersionPhysical}
\lim_{k\to 0^{+}}\omega_{hs}(q)
= \pm \frac{1}{\sqrt{2}}\sqrt{1+\beta} k
-i \frac{1}{6}\frac{r_h}{2^{1/3}} \sqrt{1+\beta} k^2 +\mathcal{O}(k^3)\, .
\end{equation}

Eq.~\eqref{eqScalarHydroDispersionPhysical} agrees exactly with the analytic sound dispersion found in~\cite{Davison2016} for $\lambda=0$. Our numerical data demonstrates that Eq.~\eqref{eqScalarHydroDispersionPhysical} is valid for all values of $\lambda$ and $\beta$.

\begin{figure}[bht]
 \includegraphics[width=\textwidth]{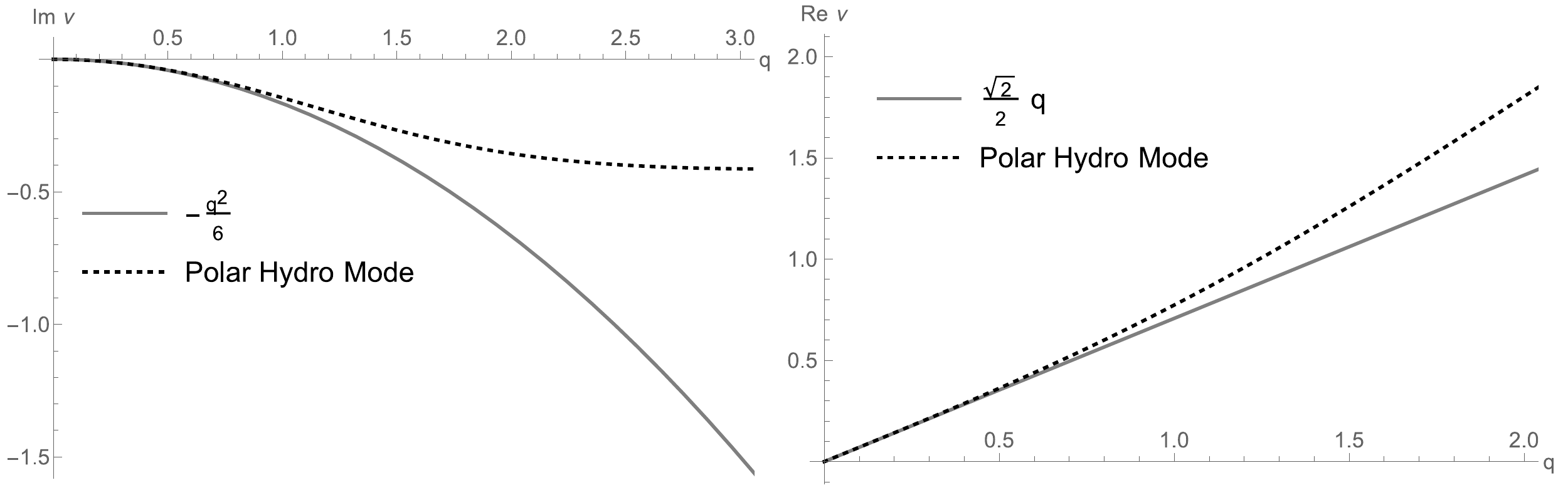}
 \caption{\label{fig:PolarHydroPlot}
 Dispersion of polar hydrodynamic modes: 
 The imaginary (left plot) and real part (right plot) of polar hydrodynamic modes is shown in dimensionless frequency variable $\nu$ versus the dimensionless momentum $q$. The exact numerical value is shown as dashed line, while the hydrodynamic approximation is displayed as solid line.}
\end{figure}

\begin{figure}[!htb]
 \includegraphics[width=\textwidth]{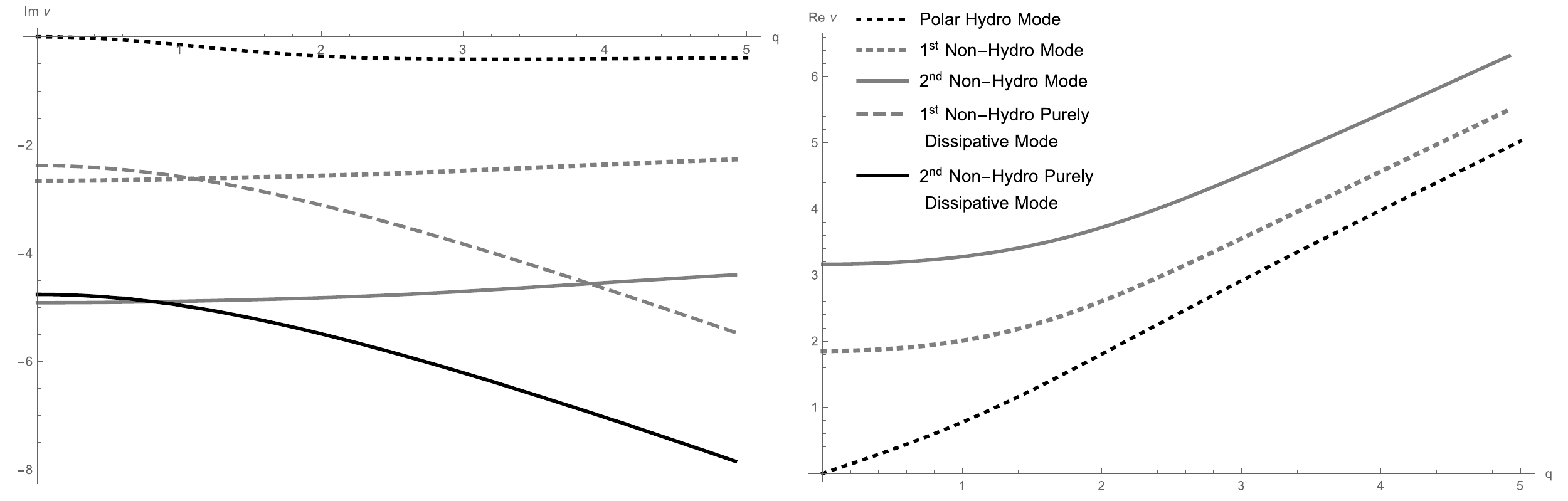}
 \caption{\label{fig:PolarNonHydroPlot}
 Dispersion of polar non-hydrodynamic QNMs: Real and imaginary part of the dimensionless QNM frequencies $\nu$ plotted versus the dimensionless momentum $q$. Note that the purely dissipative modes are only found if and only if $\lambda \neq 0$.}
\end{figure}

Unlike the axial hydrodynamic diffusion QNM, the polar hydrodynamic QNM frequencies have a non-zero real part, see Fig.~\ref{fig:PolarHydroPlot}. The mode is propagating with a speed $v_s=1/\sqrt{2}$. This value is identical to the conformal speed of sound, $1/\sqrt{d-1}$, in a relativistic $d$-dimensional field theory with two spatial dimensions~\cite{Herzog:2003ke}. The imaginary part in Eq.~\eqref{eqScalarHydroDispersion} contains the sound attenuation coefficient $\Gamma=1/6$. This is also consistent with the known relativistic formula $\Gamma=\frac{d-2}{d-1}\frac{\eta}{\epsilon+P}$~\cite{Herzog:2003ke} when factors of the spin-2 sound velocity $\sqrt{1+\beta}$ are re-instated, as already mentioned in~\cite{Davison2016}. 
Up to a rescaling of frequency with the speed factor $\sqrt{1+\beta}$, our system has the same dispersion relation as a relativistic theory, see Eq.~\eqref{eq:polarQNMEquality}, regardless of the value for $\lambda$. Hence, as expected, our fluid, and in particular the sound attenuation $\Gamma$, do not receive the corrections computed in~\cite{deBoer:2017abi}. It is noteworthy, that the sound modes do not grow at large momenta, $q>3$, see Fig.~\ref{fig:PolarHydroPlot} and~\ref{fig:PolarNonHydroPlot}. The latter figure in particular shows that a potential cross-over between imaginary parts of the hydrodynamic sound mode (Polar Hydro Mode) and the lowest non-hydrodynamic mode (1st Non-Hydro Mode) would have to occur at large momentum, $q\gg 5$.

In table~\ref{tab:smallQPolar}, we present a parametrization of the dispersion relations of the lowest 14 QNMs at small momentum $q<1$. 
It is suspicious that the higher khronon modes are approximately integer multiples of the lowest khronon frequency, e.g. at vanishing momentum 
\begin{equation}\label{eq:khrononModeEq}
    \nu_{khronon} \approx -i\, 2.381 \, n \approx -i\, \frac{3}{2^{1/3}} \, n \, , \qquad n=0, 1, 2, 3, \dots \, .
\end{equation}
This point is discussed in Sec.~\ref{sec:khrononModes}.
\begin{table}[]
\centering
\begin{tabular}{|c|c|c|c|c|}
\hline 
 $\nu _0$ & $\nu _1$ & $\nu _2$ & $\nu _3$ & $\nu _4$ \\
 \hline \hline
 $0.+0.00006 i$ & $\mp 0.70163$ & $\mp 0.07719-0.16993 i$ & $\pm 0.14775$ & $\mp 0.10496+0.02857 i$ \\
 \hline
 $0.-2.381 i$ & $0.$ & $0.$ & $0.$ & $0.$ \\
 \hline
 $1.8495  -2.66401 i$ & $0.$ & $0.1374  +0.04113 i$ & $0.$ & $0.02127  -0.01031 i$ \\
 \hline
 $-1.84948-2.66401 i$ & $0.$ & $-0.13769+0.04113 i$ & $0.$ & $-0.021-0.01031 i$ \\
 \hline
 $0.  -4.7619 i$ & $0.$ & $0.  -0.20987 i$ & $0.$ & $0.  +0.00993 i$ \\
 \hline
 $3.16133  -4.91639 i$ & $0.$ & $0.10448+0.02638 i$ & $0.$ & $0.01176  +0.0062 i$ \\
 \hline
 $-3.16137-4.91639 i$ & $0.$ & $-0.10406+0.02638 i$ & $0.$ & $-0.0122+0.0062 i$ \\
 \hline
 $4.46443  -7.16784 i$ & $0.$ & $0.08739  +0.02837 i$ & $0.$ & $0.00831  -0.01216 i$ \\
 \hline
 $-4.46445-7.16784 i$ & $0.$ & $-0.08719+0.02837 i$ & $0.$ & $-0.00852-0.01216 i$ \\
 \hline
 $5.76501  -9.41813 i$ & $0.$ & $0.07778  +0.01646 i$ & $0.$ & $0.00564  +0.01621 i$ \\
 \hline
 $-5.76504-9.41822 i$ & $0.$ & $-0.0776+0.02054 i$ & $0.$ & $-0.00542-0.00365 i$ \\
 \hline
 $7.06506  -11.6683 i$ & $0.$ & $0.06959  +0.01718 i$ & $0.$ & $0.00521  +0.0028 i$ \\
 \hline
 $-7.06509-11.6683 i$ & $0.$ & $-0.06951+0.01718 i$ & $0.$ & $-0.00486+0.0028 i$ \\
 \hline
\end{tabular}
\caption{\label{tab:smallQPolar}
Shown are the expansion coefficients for small momentum $q$ polar modes where $\nu(q) \approx \sum_{m=0}^{4} q^m \nu_m$. The two hydrodynamic sound modes are collected in the first entry. Empirically we find that the purely imaginary khronon modes are integer multiples of the lowest khronon frequency, regardless of the value of the momentum $q$, e.g. integer multiples of ``$\nu = -2.381 i$'' at $q=0$.}
\end{table}
%

\subsection{Large momentum dispersion relations (eikonal limit)} \label{sec:largeMomentumLimit}
\begin{figure}[h]
\begin{center}
 \includegraphics[width=0.95\textwidth]{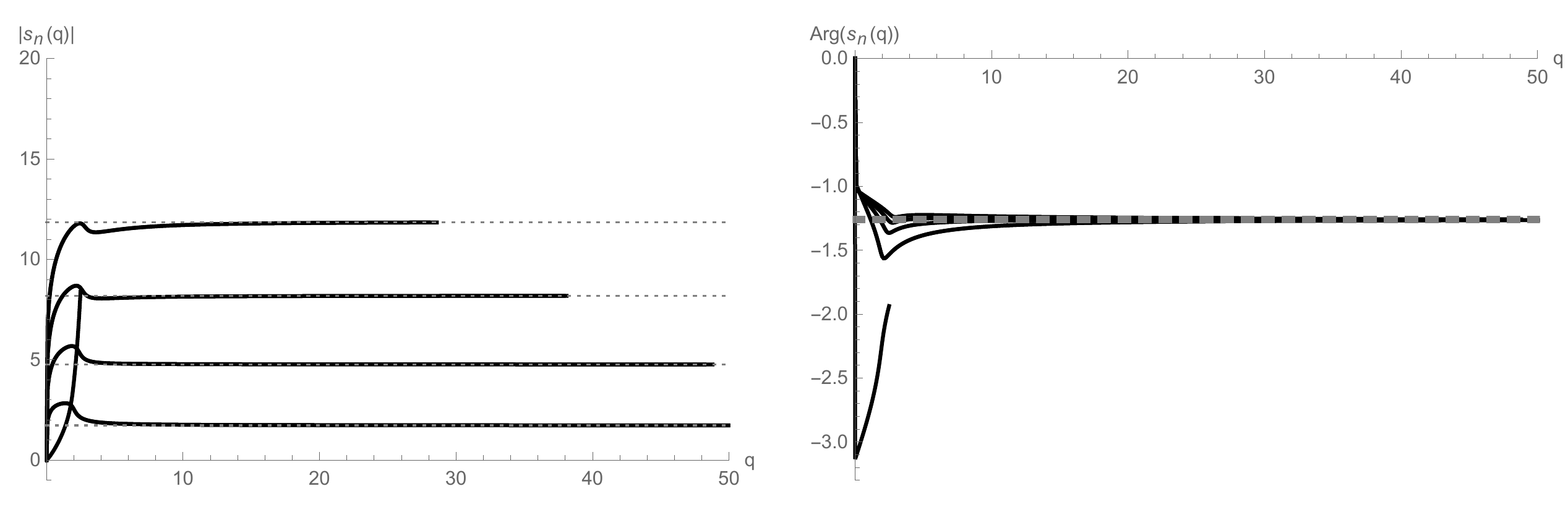}
 \caption{\label{fig:sVPlot}
 The coefficient $s_n$ are extracted from the dispersion relation of the $n$th axial mode matching it to Eq.~\eqref{eq:largeMomentumDispersion}. 
 {\it Left plot:} At large momenta, each coefficient $s_n$ approaches a distinct constant value, as predicted~\cite{Fuini2016}/ 
 The magnitudes of these constants are  $s_1\approx1.73$, 
 $s_2\approx4.76$, 
 $s_3\approx8.18$, and 
 $s_4\approx11.8$. However, the $s_{\text{diffusion}}$ of the diffusion mode (only shown for small momenta) does not seem to approach a constant value. 
 {\it Right plot:} The arguments of $s_n$ all approach $Arg(s_n)\approx-1.27$. For the diffusion mode, our data is not reliable at larger momenta as it dives deep into the complex frequency plane.
 }
\end{center}
\end{figure}
\begin{figure}[h]
\begin{center}
 \includegraphics[width=0.95\textwidth]{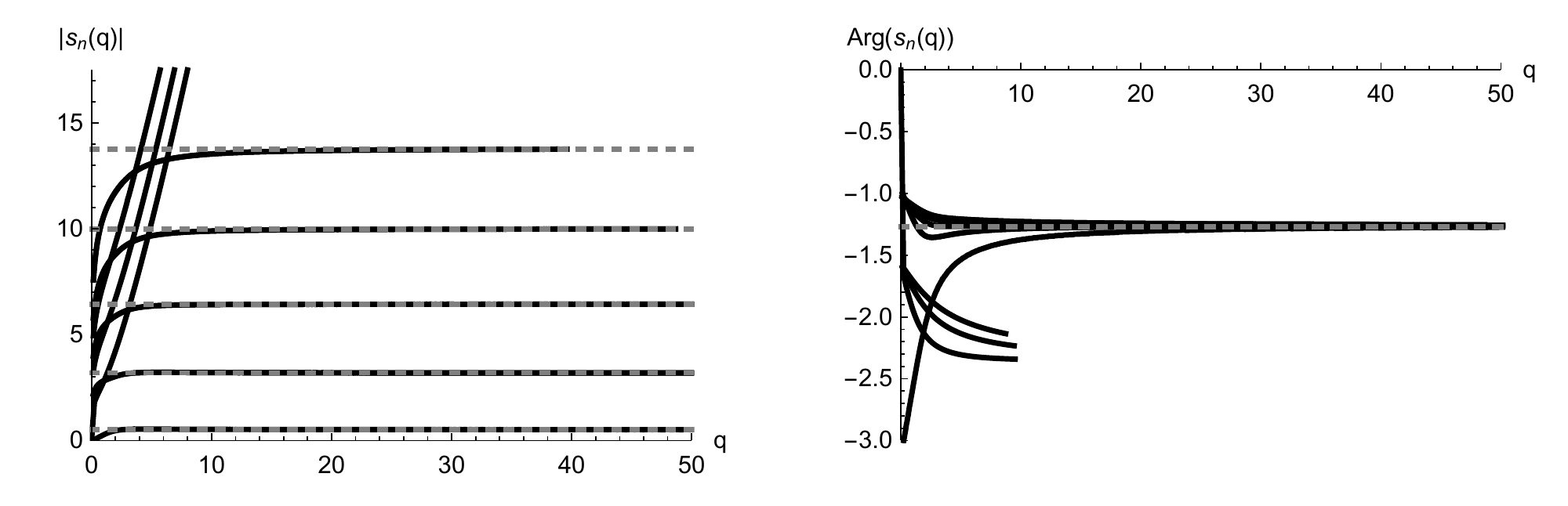}
 \caption{\label{fig:sSPlot}
 The coefficient $s_n$ is extracted from the dispersion relation of the $n$th polar mode matching it to Eq.~\eqref{eq:largeMomentumDispersion}. 
 {\it Left plot:} The magnitude of $s$ seems to asymptote to $s_1\approx0.515$, $s_2\approx3.19$, $s_3\approx6.43$, $s_4\approx9.98$, and $s_5\approx13.76$. 
 The two lines not asymptoting to any constant value belong to the two purely imaginary (khronon) modes. 
 {\it Right plot:} The arguments of $s_n$ associated with the metric all approach $Arg(s_n)\approx -1.27$.
 While the arguments of the purely imaginary modes seem to approach $Arg(s_n)\approx -\pi$.
 }
\end{center}
\end{figure}
The QNMs at large momentum are rather difficult to find. 
The amount of computation time used by the pseudospectral method code increases as momentum increases. 
At large momentum, $q\gg 1$, for both the axial and polar modes, the real part of the non-hydrodynamic frequencies are to leading order linear in the momentum.\footnote{For the purely dissipative polar non-hydro modes, the real part obviously vanishes.} 
This tendency is already seen in Fig.~\ref{fig:AxialNonHydroPlot} and Fig.~\ref{fig:PolarNonHydroPlot}, 
and we confirm numerically $Re(\nu)\approx q$. 
It was shown in \cite{Festuccia:2008zx} that QNMs of a scalar field in $AdS_4$ Einstein Gravity at large momentum $q$ at higher mode number $n$ take the form 
\begin{equation} \label{eq:largeMomentumDispersion}
\nu(q) \approx q+s_{n} ~ q^{-1/5} \qquad (q\gg 1)\, ,
\end{equation}
where $s_{n}$ is the $q^{-1/5}$ coefficient for the $n$th mode, with a phase $Arg(s_n)=\pm\pi 2/5$.  
Eq.~\eqref{eq:largeMomentumDispersion} was numerically shown to be also true for metric QNMs in $AdS_4$~\cite{Morgan:2009vg}, approximately even at lower mode numbers $n=1,2$. Analytically, Eq.~\eqref{eq:largeMomentumDispersion} was shown for $AdS_5$ metric QNMs in~\cite{Fuini2016}. 
The linear behavior $\nu\sim q$ indicates a light-like propagation and is not surprising in a relativistic theory, or in our case in a theory related to a relativistic one by mere scaling factors of the frequency and momentum, given in Eq.~\eqref{eqHorMasterEOMVars}. More interesting is the universal correction at order $q^{-1/5}$. For smaller values of $q$, we expect its coefficient to change with momentum, i.e. $s_n(q)$, but at large momentum, we expect it to asymptote to a constant independent of momentum $\lim\limits_{q\to \infty}s_n(q)\to s_n$. 
Indeed, this is what we find, as seen in Fig.~\ref{fig:sVPlot} for axial modes. All non-hydrodynamic axial QNMs approach each a value $s_n$, labeled by the mode number $n=1,2,3,4$. Their phases approach a common value of $Arg(s_n)\approx -1.27$, which is approximately $-2 \pi /5$ as expected. 
For comparison, we also show the coefficient $s_{\text{diffusion}}$ of the hydrodynamic diffusion mode, which does not approach any constant value at large momentum.  

For the polar modes, Fig.~\ref{fig:sSPlot}, there are two types of behavior. First, there is the set of modes associated with the metric perturbations, which behave according to Eq.~\eqref{eq:largeMomentumDispersion}, approaching constant values $s_n$ and a common phase $Arg(s_n)\approx -1.27$. 
Second, there are the purely imaginary khronon modes associated with the scalar field, which appear to be damped with a power different from $q^{-1/5}$, as indicated by the two trajectories not asymptoting to any constant at large momenta in Fig.~\ref{fig:sSPlot}. Their phase appears to approach $-\pi$, which indicates a negative sign for the subleading correction to the linear behavior. 
A WKB analysis similar to~\cite{Festuccia:2008zx,Fuini2016} may yield an analytic expression for the behavior of these khronon modes at large momentum. 
However, the equations of motion contain many terms and are coupled to each other, so analytically it is difficult to perform a WKB analysis. 
For now, we observe that the khronon modes do not display the large momentum behavior expected from QNMs of metric or scalar fields with higher mode numbers $n$. They rather resemble the behavior of the hydrodynamic diffusion mode shown in Fig.~\ref{fig:sVPlot}. 

\subsection{``Semi-\AE ther'' field QNMs}\label{sec:semiAether}
As an interesting aside, while analyzing the polar and axial equations of motion, we considered two additional types of perturbations of the \AE ther field, which preserve its time-like unit normality but not its hypersurface orthogonal condition:
\begin{align}
u^p_\mu(x^\sigma) &= u_\mu(x^\sigma) + \epsilon~t_\mu(x^\sigma) \label{eqAetherAditionalVectorPerts1} \, , \\
&= u_\mu(x^\sigma) + \epsilon~\partial_\mu T(x^\sigma)\label{eqAetherAditionalVectorPerts2} \, .
\end{align}
 One could claim that (\ref{eqAetherAditionalVectorPerts2}) is more ``correct'' than (\ref{eqAetherAditionalVectorPerts1}), because (\ref{eqAetherAditionalVectorPerts2}) has the correct number of degrees of freedom and preserves the hypersurface orthogonality. 
 In addition to introducing \eqref{eqAetherAditionalVectorPerts1} and \eqref{eqAetherAditionalVectorPerts2},  we also have to include the time-like unit constraint in the Einstein-\AE ther action Eq.~(\ref{eqAetherAction}), we do so by adding it with a Lagrange multiplier $\lambda_{\text{\AE}}(u^2+1)$. Then, we find $\lambda_{\text{\AE}}$ must be set to 
\begin{equation}\label{eqLambdaAetherAdditionPert}
\lambda_{\text{\AE}}= 3c_3 \left(\frac{2 c_3 r^6}{\left(c_3-1\right) r_h^6}+1\right)\, ,
\end{equation}
 in order to satisfy the unit constraint on the $u^p_\mu(x^\sigma)$ field. 
Conveniently (\ref{eqLambdaAetherAdditionPert}) works for both perturbations, \eqref{eqAetherAditionalVectorPerts1} and \eqref{eqAetherAditionalVectorPerts2}. We can derive a new set of equations of motion generated by these perturbations \eqref{eqAetherAditionalVectorPerts1}, \eqref{eqAetherAditionalVectorPerts2}. Applying a Fourier transformation and utilizing pseudospectral methods, we have two sets of axial and polar QNMs. 

In both the polar and axial sectors, the semi-\AE ther QNMs found with the new \AE ther field perturbations~\eqref{eqAetherAditionalVectorPerts1} and \eqref{eqAetherAditionalVectorPerts2} are numerically indistinguishable from those QNMs found using Eq.~\eqref{eqPerturbedMetric1}. 
This suggests that the requirement of hypersurface orthogonality does not change the QNMs in our system at hand. 

It should be noted that we find the metric fluctuation QNMs (coupled to the scalar khronon) to converge on ten significant figures at the grid size we work with, $N_{grid}=80,100$, see Fig.~\ref{fig:AxialCPPlot} and Fig.~\ref{fig:AxialNonHydroPlot}).
At the same grid size, the purely imaginary khronon modes were found to only converge on four significant figures in the case of the (\ref{eqAetherAditionalVectorPerts1}). 
%

\subsection{Khronon modes}\label{sec:khrononModes}
In this section, we discuss the {\it khronon modes}. In particular we discuss the question if the khronon modes we find are fake modes or true QNMs. As discussed before, the khronon modes are those modes in the polar sector of Einstein-\AE ther Theory, which have purely imaginary (quasi)eigenfrequencies and are non-hydrodynamic taking on a nonzero frequency value at vanishing momentum. The khronon fluctuation, $\tilde{\chi}$ couples to the other six fluctuations in the polar sector, $\tilde{h}_{xx}$, $\tilde{h}_{yy}$, $\tilde{h}_{yt}$, $\tilde{h}_{yr}$, $\tilde{h}_{rr}$, $\tilde{h}_{tt}$, and $\tilde{h}_{rt}$. At vanishing couplings $\lambda=0$ and $\alpha=0$, it is possible to analytically map the polar fluctuations to the corresponding Schwarzschild-$AdS_4$ fluctuations in Einstein Gravity using a field redefinition~\cite{Davison2016}. At nonzero $\lambda$, we find no way of decoupling the system of linear differential equations analytically. 

However, when solving the coupled system with pseudospectral methods as a generalized eigenvalue problem, we find that forcing the khronon fluctuation to vanish, $\tilde{\chi}=0$, does not affect the values of the other polar QNM frequencies (up to numerial errors). In turn, when forcing the metric fluctuations to vanish $\tilde{h}_{xx}=\tilde{h}_{yy}=\tilde{h}_{yt}=\tilde{h}_{yr}=\tilde{h}_{rr}=\tilde{h}_{tt}=\tilde{h}_{rt}=0$, we find the khronon eigenfrequencies unaffected (up to numerical errors).
\footnote{\label{foot:inconsistent} Note that this choice of vanishing metric fluctuations is inconsistent. According to the Einstein equations, also the khronon has to vanish in this case.
Potentially the khronon field can be shown to decouple after a field transformation or use of master fields~\cite{Kodama:2000fa}.}
Hence, we conclude that the khronon {\it numerically decouples} from the metric fluctuations in the polar sector. 

We furthermore observe that the khronon eigenfrequencies, or khronon modes, assume purely imaginary values, which are integer multiples of the lowest khronon mode, $\nu_{khronon} = - i\, 2.381 \, n$ with $n=0, 1, 2, 3, \dots$, as stated in Eq.~\eqref{eq:khrononModeEq}. In fact, if we change our frequency normalization by a factor to $\hat\nu = \nu 2^{1/3}$, then $\hat\nu_{khronon} = - i\, 3 \, n$. Such integer value solutions are known from various analytical solutions for quasinormal mode frequencies~\cite{Starinets:2002br}. However, such behavior is also known from various fake modes~\cite{Janiszewski:2015ura}. The latter are normally revealed because they either do not converge to any value as accuracy is improved (grid size is increased), and such fake modes normally do not change their frequency with changing momentum. However, we find that the khronon modes converge to fixed frequencies, although not as quickly as the metric QNMs. The khronon modes converge to four significant figures while the metric QNMs converge to ten at a grid size of $N_{grid}=100$. Furthermore, the khronon modes do move with momentum, as illustrated in Fig.~\ref{fig:sSPlot} and Fig.~\ref{fig:PolarNonHydroPlot}. The latter figure indicates a quadratic rise at small momentum $q<2$ and a linear rise thereafter.
Hence, the convergence and momentum dependence indicate that the khronon modes are not fake modes. 

The khronon is a scalar field, but its equation of motion is not written in the standard Klein-Gordon form. This comes from the fact, that the khronon enters the Einstein \AE ther action~\eqref{eqAetherAction} through dynamical terms for the vector $u_\mu \propto \partial_\mu \phi$, which are quadratic in derivatives on $u_\mu$. This leads to a fourth order equation of motion for $\phi$ or its fluctuation $\tilde{\chi}$, see Eq.~\eqref{eqPerturbedMetric1}. So it is worthwhile analyzing this fourth order equation separately by forcing the metric fluctuations to vanish, see footnote~\ref{foot:inconsistent}. Our near-horizon analysis reveals that this fourth order equation has a regular singular point at the universal horizon $r=r_h=1$. There are four indicial exponents for the khronon fluctuation near the horizon $\tilde{\chi}\approx \chi_0 (1-r)^\alpha$, which are all of the form
\begin{equation}
    \alpha = i\frac{\hat\nu}{3} + \mathfrak{f}(q) \, 
\end{equation}
with a momentum dependent real-valued function $\mathfrak{f}(q)$, and we recall that $\hat\nu=2^{1/3} \nu$. It is a novelty for the indicial exponent to depend on momentum as this is not the case for any QNM equation as far as we know. This momentum dependence can be traced back to the equation being fourth order. In general, $f(q)$ is rather complicated. Let us consider first the case of vanishing momentum, $q=0$. Then the four indicial exponents simplify to 
\begin{equation}\label{eq:indicialExponentsKhronon}
    \alpha(q=0) = -3+i\frac{\hat\nu}{3},\, -2 +i\frac{\hat\nu}{3},\, -1 +i\frac{\hat\nu}{3},\, i\frac{\hat\nu}{3} \, .
\end{equation}
At this point, we recall that the equations of motion are written in terms of Eddington-Finkelstein coordinates, such that ingoing modes at the horizon are regular, others are singular. None of the modes in Eq.~\eqref{eq:indicialExponentsKhronon} is regular at the horizon, except for special values of $\nu$. Those special values are $\hat\nu= - i\, 3\, n$ with $n=0,1,2,3, \dots$. Generalizing the momentum to $q>0$, we find that regular modes appear at frequency values 
\begin{equation} \label{eq:regularKhrononModes}
    \hat\nu = - i\, 3\, (n-\mathfrak{f}(q)) \, , \qquad n=0,\, 1,\, 2,\, 3,\, \dots . 
\end{equation}
These are the khronon modes found by our pseudospectral method when solving the generalized eigenvalue problem. Eq.~\eqref{eq:regularKhrononModes} explains the momentum dependence discussed above. At $q=0$, Eq.~\eqref{eq:regularKhrononModes} also explains the observation that khronon mode frequencies are integer multiples of 3 when written in terms of $\hat\nu$. The numerical data indicates that this property also holds at $q\neq 0$, which implies that $\mathfrak{f}(q)\propto n$. Now the question remains, if these khronon modes are to be regarded as true QNMs or as fake modes.\footnote{It is a logical possibility that there exist ingoing solutions for the khronon, which are not regular in the coordinates and field definitions we have chosen here. However, it is not clear to us how to find such modes.}

One defining property of a QNM is that it vanishes at the AdS-boundary. If the khronon modes are QNMs, then they have to vanish at the AdS-boundary. This can be checked by calculating the eigenvectors associated with the purely imaginary khronon frequencies in question. This analysis is technically very difficult in the full system, because the relevant matrix in the eigenvalue problem is not invertible. Since the khronon modes and the metric QNMs seem to numerically decouple, we hence restrict our eigenvector analysis to the case in which we force all metric fluctuations to vanish as above. In that case, the matrix is invertible, and we confirm that all khronon modes assume non-trivial values along the radial direction and all vanish at the AdS-boundary.\footnote{As an interesting aside, in this case we also observe that at larger momenta $q>3/2$, the khronon spectrum contains both purely imaginary and also propagating modes with real and imaginary part to their QNM frequencies. This seems interesting in light of the observation that only one of these two types appeared~\cite{Sybesma2015} at vanishing momentum, depending on the dynamical scaling $z$ and the number of dimensions. However, this behavior is not observed when the metric is allowed to fluctuate. Hence, this appears to be an artifact of the artificial decoupling.}

As a further test, we consider the large momentum $q\gg 1$ limit, also called eikonal limit. Results for the two purely imaginary khronon modes were already discussed in Sec.~\ref{sec:largeMomentumLimit}, and presented in Fig.~\ref{fig:sSPlot}. The observed phase is $Arg(s_n)\approx -3\pi/4$ and the subleading correction is not of order $q^{-1/5}$. This behavior is neither that of a scalar nor that of a metric fluctuation. However, that may not be too surprising because the khronon does not satisfy a simple linearized scalar equation of motion of second order. It rather satisfies a fourth order equation and one should probably conduct a WKB analysis for the vector $u_\mu$ and compare its numerical behavior with what is expected from that analysis. However, such a treatment is beyond the scope of this work. 

Our large momentum analysis ends up being inconclusive. 
However, the khronon modes (at least when decoupled from the metric fluctuations) satisfy the two defining relations of a QNM: they vanish at the AdS-boundary, and are ingoing at the horizon.
Based on this, we decide to interpret the khronon modes as true QNMs which are part of the polar sector of the theory. We speculate that our limitation of $\alpha=0$ is forcing part of the khronon dynamics to vanish. This is plausible because the term in the \AE ther action~\eqref{eqAetherAction} set to zero by $\alpha=0$ (equivalent to $c_4=0$) is essentially quadratic in a time derivative of the vector $u_\mu$. We speculate that $\alpha\neq 0$ would allow the khronon modes to propagate. In that case, however, the analytic background solution is not valid anymore and one has to work with numerical background solutions~\cite{Janiszewski2015}, which is left for future work.

\section{Summary \& conclusions}\label{sec:conclusions}
%
In this paper we have calculated non-relativistic gravitational QNMs on an analytically known asymptotically $AdS_4$ black brane solution, Eq.~\eqref{eqADSHorBlackGroundMetric} and~\eqref{eqAetherMatrix}, with one of the three coupling constants vanishing, $\alpha=0$~\cite{Janiszewski2015}. The theory is comprised of two sectors, the parity even polar sector, and the parity odd axial sector. Each sector consists of gravitational fields which travel at a certain speed, either the spin-0 or the spin-2 speed. Correspondingly, the relevant horizon for the axial sector is the spin-2 sound horizon experienced by the spin-2 graviton. While the relevant horizon for the polar sector is the spin-0 sound horizon experienced by the spin-0 graviton. 
%
We presented QNMs up to mode number $n=5$ for both sectors over a large range of Ho\v{r}ava couplings $\lambda$, $\beta$, and including large momentum up to $q=50$. 
Our results are summarized in Fig.~\ref{fig:ParameterPolarPlots} and Fig.~\ref{fig:ParameterAxialPlots} for the polar and axial sector, respectively.
Equations of motion, and QNM data is collected in four ancillary files.

In this work, we have shown numerically that all Einstein Gravity QNMs are contained in the set of QNMs of Ho\v{r}ava Gravity for any value of $\lambda, \beta$ at $\alpha=0$, when expressed in appropriate units. At $\lambda\neq 0$, Ho\v{r}ava Gravity has an additional set of purely imaginary QNMs, the khronon modes. The khronon modes seem to numerically decouple from the metric modes and we conjecture an analytic dispersion relation, Eq.~\eqref{eq:regularKhrononModes}, 
\begin{equation}
    \omega_{\text{khronon}} = - i\frac{\sqrt{1+\beta}}{r_h} 3\, (n-\mathfrak{f}(q))\, ,\qquad n=1,\, 2,\, 3,\, \dots \, .
\end{equation}
Furthermore, we conjecture an analytic relation between the QNMs of Einstein Gravity and all QNMs of Ho\v{r}ava Gravity at arbitrary $\lambda$, $\beta$, and at $\alpha=0$, except the khronon modes:
\begin{equation}
    \omega_{\text{Ho\v{r}ava}} = \frac{r_{\text{sound}}}{r_{\text{Schwarzschild}}}\sqrt{1+\beta} \, \omega_{\text{Einstein}} \, ,
\end{equation}
where $r_{\text{Schwarzschild}}$ is the Schwarzschild horizon of a black brane in Einstein Gravity, and $r_{\text{sound}}$ is the sound horizon relevant for each sector of QNMs in the analytic Ho\v{r}ava Gravity black brane solution, Eq.~\eqref{eqAetherMatrix}. That is the universal horizon $r_h$ in the polar sector, and the spin-2 sound horizon $r_h/2^{1/3}$ in the axial sector. 
In other words, the QNM frequencies in Einstein Gravity and in the analytic Ho\v{r}ava black brane solution, measured in units of the respective horizon, are equal to each other except for a factor of $\sqrt{1+\beta}$. 
%

In the axial sector, at any value of $\lambda$ and $\beta$, there is one hydrodynamic diffusion mode and a set of propagating (not overdamped) non-hydrodynamic QNMs, see Fig.~\ref{fig:AxialCPPlot}. 
The absence of overdamped (purely imaginary modes) in this sector is in agreement with the claim from~\cite{Sybesma2015,Gursoy:2016tgf}. 
The hydrodynamic diffusion mode starts out having quadratic dispersion at small momentum in agreement with the analytic prediction, Eq.~\eqref{eq:VectorHydroDispersion} and Fig.~\ref{fig:AxialHydroPlot}. However, then it increases faster in magnitude around $q=1$. Around $q=2$, it is damped more than the lowest non-hydrodynamic mode (with mode number $n=1$). 
This has been observed before in relativistic theories~\cite{Kaminski:2009ce}. 
At large momentum, $q\gg 1$, the non-hydrodynamic modes dominate the system since their damping decreases and they are long lived as seen from Fig.~\ref{fig:AxialNonHydroPlot}. 
Dispersion relations for the lowest 9 axial QNMs are parametrized to fourth order in momentum in table~\ref{tab:smallQAxial}.
%

In the polar sector, we distinguish two cases, $\lambda=0$ and $\lambda\neq 0$, see Fig.~\ref{fig:PolarCPPlot}. If $\lambda=0$, then there are no modes associated with the khronon field, only those associated with the metric. Those are two hydrodynamic sound modes and a set of non-hydrodynamic QNMs. The sound modes at small momentum $q<1$, see Fig.~\ref{fig:PolarHydroPlot}, agree with the linear propagation and quadratic damping in Eq.~\eqref{eqScalarHydroDispersionPhysical}, which was derived only at $\lambda=0$. Our analysis shows that this equation holds also at $\lambda\neq 0$. Again, at large momentum, $q\gg1$, the system is likely dominated by the non-hydrodynamic modes, because again their damping decreases, as seen in Fig.~\ref{fig:PolarNonHydroPlot}. 
Although, this cross-over probably occurs at a much larger momentum than in the axial sector. This is because the polar hydro modes (sound modes) seem to asymptote to a constant value between $0$ and $-i$ for large momentum.
In addition to that, in the other case, $\lambda\neq 0$, purely imaginary khronon modes are present. In that case, our QNM spectrum contains both overdamped and non-overdamped modes. The overdamped modes are associated with the scalar khronon field fluctuation, while the non-overdamped modes are associated with the metric fluctuations. This is interesting in the context of the claim that only one type, namely overdamped or non-overdamped modes should appear at a given combination of dynamical exponent $z$ and number of dimensions $d$~\cite{Sybesma2015,Gursoy:2016tgf}. The latter works consider cases in which a massive scalar probe field does not couple to the metric fluctuations. Hence, it would be interesting in which form the claim needs to be generalized to coupled systems like the one we have studied here.   
Dispersion relations for the lowest 14 QNMs are parametrized to fourth order in momentum in table~\ref{tab:smallQPolar}.
%

We have also performed a large momentum analysis (eikonal limit) and found to match the analytic expectation based on~\cite{Morgan:2009vg,Festuccia:2008zx,Fuini2016}, see Fig.~\ref{fig:sVPlot} and~\ref{fig:sSPlot}. At large momentum, $q\approx 50$, all our metric (non-overdamped) QNMs follow the relation $\nu(q)\approx q+s_n q^{-1/5}$, asymptoting to a constant magnitude for $s_n$ and with a universal phase $Arg(s_n)\approx -\pi 2/5$. 
Our overdamped modes, the khronon modes, however, do not show the large momentum dispersion expected from either a scalar or a metric perturbation QNM. As seen in Fig.~\ref{fig:sVPlot} and~\ref{fig:sSPlot}, their $s_n$ values do not asymptote to constants and their phase is not $\pm \pi 2/5$. 
%

It is interesting to speculate about why the khronon modes decouple from the other modes in the polar sector. The limit of infinite speed is likely the reason for this. A nonzero $\alpha$ leads to finite speed and a horizon for the khronon which will be different from the universal horizon. This case then allows for time derivatives of the khronon in the actions~\eqref{eqActionHor} and~\eqref{eqAetherAction}. Moreover, also the metric modes in the polar sector will travel at a finite speed, which can be distinct from the khronon speed, depending on the Ho\v{r}ava couplings~\cite{Jacobson2004}. It will be interesting to see the dynamics and interplay of fields in the polar sector in this case. 

From a physical perspective, it is remarkable, that the relativistic relation between shear viscosity and sound attenuation, $\Gamma \propto \eta$, still holds in this solution of Ho\v{r}ava Gravity. In the latter, gravitational fields (in the axial sector) giving rise to shear modes travel at a different speed than those (in the polar sector) giving rise to sound modes. So one could generally have expected that physical quantities from the polar sector have nothing to do with the axial sector. It would be interesting to check this relation at nonzero $\alpha$.
%

In passing, we have verified the equivalence of axial QNMs derived from hypersurface orthogonal Einstein-\AE ther Theory and those derived from Ho\v{r}ava Gravity. 
%
Remarkably, the hypersurface orthogonality does not influence the QNMs in our system, as our semi-\AE ther results indicate, see Sec.~\ref{sec:semiAether}.

In summary, an obvious, though numerically challenging extension of this work would be the calculation of QNMs for Ho\v{r}ava Gravity with nonzero $\alpha$ coupling, or equivalently Einstein-\AE ther theory with nonzero $c_4$. It will especially be interesting to see the full dynamics of the khronon field unfold. It is expected that QNMs will shift compared to the case $\alpha=0$, and that the polar sector should be truly coupled, with one common set of QNMs. In that setting, one should check how the prediction of purely imaginary modes at $d\le z+1$ needs to be modified for a khronon fluctuations coupling to metric fluctuations. 
A technical improvement may simplify this computation: Possibly gauge invariance could be used to define master fields, reducing the number of fields and field equations that need to be solved.

Relativistic hydrodynamics has been systematically constructed and restricted as an effective field theory over the past years. While Lorentz covariance serves as a fundamental construction principle in that case, it was less clear how to construct non-relativistic hydrodynamics systematically. One way is to start from a relativistic hydrodynamic description, e.g.~\cite{Jensen:2011xb}, and then take a non-relativistic limit sending the speed of light to infinity~\cite{Kaminski:2013gca}, where the choice of the hydrodynamic frame is important~\cite{Jensen:2014wha}. A second way is to identify the non-relativistic data structures directly, as is done in the context of Newton-Cartan geometry~\cite{Son:2013rqa,Jensen:2014aia,Jensen:2014ama}. It has been shown that dynamical Newton-Cartan geometry gives rise to Ho\v{r}ava Gravity~\cite{Hartong:2015zia}. Thus, it would be interesting to use Ho\v{r}ava Gravity as a framework for testing explicitly the proposals for non-relativistic hydrodynamics mentioned above. This may reveal inconsistencies or lead to the discovery of neglected effects. 

\acknowledgments
MK thanks C.~Uhlemann for helpful correspondence. 
We thank R.~Davison for helpful comments on the manuscript. 
This work was supported, in part, by the U.S.~Department of Energy grant DE-SC-0012447.

\appendix

\section{Tabulated data}\label{abTabulatedData}
Exemplary QNM values for comparison can be found in tables~\ref{tableAetherAxialData} and~\ref{tableAetherPolar}. Much more extensive data has been included with this submission as ancillary files. 
\begin{table}[]
\centering
 \begin{tabular}{||c c c||} 
 \hline
 Momentum ($q$) & Frequency ($\nu$) & $n^{th} Mode$ \\ [0.5ex] 
 \hline\hline
 $0.1$ & $-0.0033361486~i$ & 0 \\
  \hline
  $2.5$ & $-6.5109332704~i$ & 0 \\
  \hline
$0.1$ & $\pm 1.8508043279-2.6634620863~i$ & 1 \\
  \hline
$2.5$ & $\pm 2.5648819977-1.7693681441~i$ & 1 \\
  \hline
$20.0$ & $\pm 20.2726721388-0.9160839493~i$ & 1 \\
  \hline
$0.1$ & $\pm 3.1623087258-4.9161394118~i$ & 2 \\
  \hline
  $2.5$ & $\pm 3.3930208613-4.2840047943~i$ & 2 \\
  \hline
  $20.0$ & $\pm 20.794701895-2.4963744671~i$ & 2 \\
  \hline
\end{tabular}
\caption{A Sample of the Einstein-\AE ther/Ho\v{r}ava Axial Quasinormal-Modes, found using pseudospectral methods.}
\label{tableAetherAxialData}
\end{table}
\begin{table}[]
\centering
 \begin{tabular}{||c c c||} 
 \hline
 Momentum ($q$) & Frequency ($\nu$) & $n^{th} Mode$ \\ [0.5ex] 
 \hline\hline
   0.25198 & $\pm 4.227043-0.006645~i$ & 0 \\
   \hline
     2.51984 & $\pm 2.144945-0.454586~i$ & 0 \\
   \hline
     7.55953 & $\pm 7.570777-7.55953~i$ & 0 \\
   \hline
     0.25198 & $\pm 2.331225-3.355934~i$ & 1 \\
   \hline
     2.51984 & $\pm 2.941333-4.227044~i$ & 1 \\
   \hline
     7.55953 & $\pm 8.473500-3.621822~i$ & 1 \\
   \hline
     0.25198 & $\pm 5.022401-7.803214~i$ & 2 \\
   \hline
     2.51984 & $\pm 5.535820-7.707342~i$ & 2 \\
   \hline
     7.55953 & $\pm 9.780225-7.023156~i$ & 2 \\
   \hline
    0.25198 & $-3.788146~i$ & 1 (Purely Dissipative ) \\
   \hline
    4.78770 & $-6.099892~i$ & 1 (Purely Dissipative ) \\
   \hline
\end{tabular}
\caption{\label{tableAetherPolar}
A Sample of the Einstein-\AE ther Polar Quasinormal-Modes, found using pseudospectral methods.}
\end{table}

\section{Convergence}\label{sec:convergence}
\begin{figure}[h]
    \centering
\includegraphics[width=\textwidth]{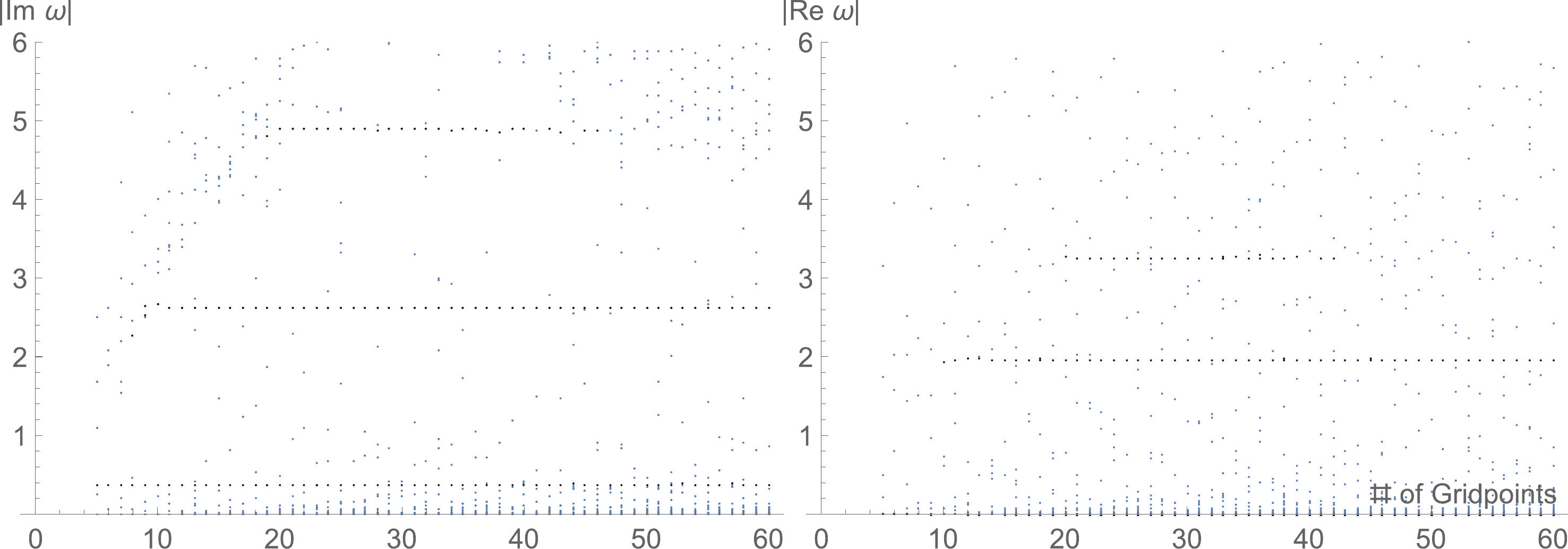}
    \caption{\label{fig:convergencePlot}
    Comparison of solutions found via the spectral method. The imaginary part (left plot) and real part (right plot) are shown as a function of the number of grid points used. Blue dots are non-convergent fake modes, black dots mark convergent quasinormal mode frequencies. 
    }
\end{figure}
Following~\cite{Jansen2017}, the pseudo-spectral method used in this paper is essentially an eigenvalue solver. Simply put, the expansion of fluctuations into polynomials of $n$th order, allows $n$ eigenvalues. The number $n$ increases with the size of the chosen grid $N_{grid}$. Nevertheless, not all of the eigenvalues found are quasinormal frequencies. Some of them are {\it fake modes}. In order to determine which eigenvalues are quasinormal modes, we compare two sets of eigenvalues, found using two different grid sizes. The frequencies that do not move more than a set distance $\Delta \omega_{cutoff}$ in the complex plane are deemed to be convergent and are called a quasinormal frequency. For example, with $k=1$ and $c_3=0$ the absolute values of the imaginary part and real part of eigenvalues are displayed in Fig.~\ref{fig:convergencePlot}. 
For modes of physical interest (black dots in the plot), we can see not much change for increasing grid size. In contrast to that, the blue dots representing fake mode eigenvalues, change appreciably. However, the number of grid points alone does not always determine convergence. For larger number of grid points used, we find that numerical errors from insufficient numerical precision in our calculation ruins convergence. So a higher numerical precision is required to find higher order modes. We use 80 and 100 grid points and a cutoff value of $\Delta \omega_{cutoff}=10^{-8}$ to test for convergence. The precision of our arithmetic is $\thicksim 10^{-45}$. For polar modes, we found some unphysical modes which did not change up to precision used with changes in momentum. 
Here, ``did not change up to precision used'' means that with the digits used, these modes seemed to not change as momentum varied. Physical modes are expected to change with momentum. 
We discard these non-varying modes due to this pathological behavior. 

\section{Khronon scalar field}\label{abEOM}
The Khronon scalar field takes the following form.
\begin{align*}
\phi(r,\nu) = & \frac{1}{6 \sqrt{1-c_3}} \bigg({(1-\sqrt{1-c_3}-c_3)}^{2/3} {(c_3-1)}^{1/3} r_h \log \big(-{(c_3-1)}^{1/3} \\ & \qquad \qquad \qquad \qquad \times {(-\sqrt{1-c_3}-c_3+1)}{(c_3-1)}^{1/3}r r_h \\ & \qquad \qquad \qquad +{\left(c_3-1\right)}^{2/3} r_h^2+{(1-\sqrt{1-c_3}-c_3)}^{2/3} r^2\big) \bigg)  \nonumber \\ & + \frac{r_h \log \left(r r_h+r_h^2+r^2\right)}{6 \sqrt{1-c_3}}-\frac{c_3 r_h \log \left(r r_h+r_h^2+r^2\right)}{6 \sqrt{1-c_3}}+\frac{c_3 r_h \log \left(r_h-r\right)}{3 \sqrt{1-c_3}}\\ &-\frac{r_h \log \left(r_h-r\right)}{3 \sqrt{1-c_3}}  \nonumber \\ &-\frac{{(1-\sqrt{1-c_3}-c_3)}^{2/3} \sqrt[3]{c_3-1} r_h \log \left(\sqrt[3]{c_3-1} r_h+\sqrt[3]{-\sqrt{1-c_3}-c_3+1} r\right)}{3 \sqrt{1-c_3}}  \nonumber \\ &-\frac{{(1-\sqrt{1-c_3}-c_3)}^{2/3} {(c_3-1)}^{1/3} r_h \tan ^{-1}\left(\sqrt{\frac{1}{3}} \left(\frac{2 {(1-\sqrt{1-c_3}-c_3)}^{1/3} r}{{(c_3-1)}^{1/3} r_h}-1\right)\right)}{\sqrt{3} \sqrt{1-c_3}}  \nonumber \\ &+\frac{r_h \tan ^{-1}\left(\frac{r_h+2 r}{\sqrt{3} r_h}\right)}{\sqrt{3} \sqrt{1-c_3}}-\frac{c_3 r_h \tan ^{-1}\left(\frac{r_h+2 r}{\sqrt{3} r_h}\right)}{\sqrt{3} \sqrt{1-c_3}} - \nu \nonumber
\end{align*}
%

%
\bibliographystyle{JHEP}
\bibliography{main}

\end{document}